\pdfoutput = 1 

\documentclass[aps,prd,twocolumn,nofootinbib,superscriptaddress,floatfix,showpacs,showkeys,preprintnumbers,
  amsmath,amssymb, aps,prd]{revtex4-2}

\usepackage{cancel}
\usepackage{multirow}

\usepackage{graphicx}
\usepackage{dcolumn}
\usepackage{bm}

\usepackage{hyperref}

\usepackage[english]{babel}
\usepackage{xcolor}
\usepackage{mathtools}
\usepackage{pdfsync}
\usepackage{amssymb}
\usepackage{amsmath}

\usepackage{slashed}

\usepackage[squaren]{SIunits}

\usepackage[active]{srcltx}

\def \beq  {\begin{equation}}
\def \eeq  {\end{equation}}

\newcommand{\Dleft}{\stackrel{\leftarrow}{D}_+}
\newcommand{\Dright}{\stackrel{\rightarrow}{D}_+}

\newcommand{\NS}{\scriptscriptstyle{\text{NS}}}
\newcommand{\sing}{\scriptscriptstyle{\text{S}}}

\def\II{\hbox{{1}\kern-.25em\hbox{l}}}

\newcommand{\Aboldcal}{\bm{\mathcal{A}}}
\newcommand{\Rboldcal}{\bm{\mathcal{R}}}
\newcommand{\Oboldcal}{\bm{\mathcal{O}}}
\newcommand{\Zboldcal}{\bm{\mathcal{Z}}}
\newcommand \widebar [1] {\overline{#1}}

\begin{document}
\allowdisplaybreaks

\preprint{DESY-25-141}

\begin{abstract}
We use conformal symmetry  
to calculate the NNLO anomalous dimension matrix (three loops) for flavor-singlet axial-vector QCD
operators for spin $N \le 8$ from a set of gauge-invariant two-point correlation functions. 
Combining this result with the recent calculation of the two-loop coefficient functions, 
we carry out the calculation of the  $\gamma\gamma^\ast\to \eta$ and  $\gamma\gamma^\ast\to \eta'$ form factors
at large momentum transfers to the NNLO accuracy in perturbative QCD.     
\end{abstract}

\title{
The  $\gamma^\ast\to \eta\gamma$ and  $\gamma^\ast\to \eta'\gamma$ form factors to NNLO accuracy in perturbative QCD}

\author{Vladimir~M.~\surname{Braun}}

\affiliation{Institut f{\"u}r Theoretische Physik, Universit{\"a}t Regensburg, D-93040 Regensburg, Germany}

\author{Konstantin~G.~\surname{Chetyrkin}}

\affiliation{Institut f\"ur Theoretische Teilchenphysik, Karlsruher Institut f\"ur Technologie, 
  D-76131, Karlsruhe, Germany}



\author{Alexander~N.~\surname{Manashov}}

\affiliation{II.~Institut f\"ur Theoretische Physik, Universit\"at Hamburg, D-22761 Hamburg, Germany}

\affiliation{Institut f{\"u}r Theoretische Physik, Universit{\"a}t Regensburg, D-93040 Regensburg, Germany}

\date{\today}
\maketitle
\section{Introduction}\label{sect_introduction1}%

The studies of the pseudoscalar $\eta$  and $\eta'$ mesons have a long history. 
In the exact flavor $SU(3)$  limit the $\eta$ meson
is part of the flavor-octet whereas the $\eta'$ is a flavor-singlet state whose properties are
intimately related to the  axial anomaly \cite{Witten:1978bc,Veneziano:1979ec}.
The $SU(3)$-breaking effects are, however, large and have a non-trivial structure. These effects
are usually described in terms of a mixing scheme that considers the physical
mesons as superpositions of fundamental flavor-singlet and flavor-octet fields in 
a low-energy large-$N_c$ chiral perturbation theory (ChPT). 
When combined with dispersion relations, this approach provides a quantitative description of a large variety of 
low energy $\eta$ and $\eta'$  production processes and decays, see, e.g., \cite{Gan:2020aco} and
references therein. In recent years, lattice QCD calculations of the relevant parameters
\cite{Bali:2021qem, Ottnad:2025zxq} are becoming increasingly precise.

A complementary and less explored area is the $\eta,\eta'$ production in hard reactions.
It is not obvious whether and to what extent the approaches based
on low energy effective field theory provide an adequate description of such processes, that 
are sensitive to  meson wave functions at small interquark separations, referred to as
light-cone distribution amplitudes (LCDAs). 
One open question in this context is that $\eta,\eta'$ mesons, in contrast to the pion, 
might contain a substantial admixture of a ``gluonium''  or ``intrinsic charm''  component 
at low scales, i.e. a sizable two-gluon or  $c\bar c$ LCDA. 
This possibility was explored in different setups~\cite{Feldmann:1998vh,Kroll:2002nt,Blechman:2004vc,Kroll:2012gsh,Agaev:2014wna},
but without a  definite conclusion.

The theoretically cleanest setup to check the applicability of the conventional state mixing concept 
for the LCDAs and also set upper limits on the ``gluonium'' LCDA is provided by two-photon processes.
The challenge is, however, to go over to largest possible momentum transfers to avoid contamination by the so-called
``end-point'' or ``Feynman mechanism'' contributions which are poorly understood. 
The BaBar collaboration measured in 2006  \cite{BaBar:2006ash} the cross section for the reaction
$e^+e^- \to  \gamma^\ast \to M \gamma$ with $M = \eta,\eta'$ at $\sqrt{s} = 10.58$~GeV, corresponding to 
the (positive) photon virtuality $q^2 = s = 112$ GeV$^2$. This cross section is related to the (absolute value squared of the)
time-like transition form factor (FF)
\begin{align}
& \hspace*{-1cm}i \int d^4 x\, e^{- i q x} \langle 0 |T\{j^{\rm em}_\mu(x)j^{\rm em}_\nu(0) \}|M(p)\rangle
\notag\\
&=  e^2 \varepsilon_{\mu\nu\alpha\beta} q^\alpha p^\beta F_{M}(Q^2)\,.
\end{align}
The result is (in MeV) 
\begin{align}
 q^2 |F_\eta(q^2)| &= 0.229 \pm 0.030 \pm 0.008\,, 
\notag\\
 q^2 |F_{\eta'}(q^2)| &= 0.251 \pm 0.019 \pm 0.008\,,
\label{BaBar-data}
\end{align}
where the first error is statistical and the second one systematic. 
The Belle II experiment will be able to decrease the errors significantly, see Section~15.7 in \cite{Belle-II:2018jsg}. In addition, some
of the parameters can nowadays be  calculated with
high precision on the lattice. The comparison of the QCD calculations including lattice input with
expected experimental results will allow one to study the structure of $\eta$, $\eta'$ mesons at short interquark
separations, encoded in the LCDAs, on a quantitative level.
This, in turn, will benefit theory studies of B decays in final states involving $\eta$ and $\eta'$  mesons. 
The transition form factor studies at time-like momentum transfers will eventually be complemented by studies of very rare
exclusive decays of electroweak gauge bosons, e.g. $Z\to\eta\gamma$ \cite{Alte:2015dpo}, in the high luminosity run at the
LHC or, later, at a future lepton collider.

Measurements of the $\gamma\gamma^\ast\eta(\eta')$ transition form factors in the  space-like region $q^2<0$ are 
also available \cite{BaBar:2011nrp}, but are restricted
to smaller values of the momentum transfer $|q^2|<35\,\text{GeV}^2$. They are still affected by sizable power-suppressed
end-point contributions, see a detailed discussion and the LCSR-based calculation of such corrections in Ref.~\cite{Agaev:2014wna},
and further references. 
In this work we will stay strictly within the leading-twist factorization framework and for this reason focus on the predictions 
for the time-like form factors where the highest possible momentum transfers can be achieved.

The principal new contribution of this paper is the calculation
of the full NNLO anomalous dimension matrix (three loops) for flavor-singlet axial-vector QCD
operators for spin $N \le 8$. This calculation is done using the technique developed in Ref.~\cite{Braun:2022byg}
which employs conformal symmetry of QCD at the Wilson-Fisher critical point in noninteger $d=4-2\epsilon$ space-time 
dimensions. In this approach 
the anomalous dimension matrix is obtained from a set of two-point correlation functions so that 
the complications due to mixing of gauge-invariant with gauge-non-invariant operators are completely avoided. 
Combining this result with the recent calculations of the two-loop coefficient functions 
\cite{Braun:2020yib,Gao:2021iqq,Braun:2021grd,Braun:2022bpn,Ji:2023xzk}, 
we obtain the  $\gamma\gamma^\ast\to \eta$ and  $\gamma\gamma^\ast\to \eta'$ form factors
at large momentum transfers to the NNLO accuracy in perturbative QCD.    
The uncertainty of these predictions as measured conventionally by the residual scale dependence is reduced to about 1\%. 

The second new element of our analysis is a consistent implementation of the contributions of charm quarks using a variable 
flavor number scheme. This approach has become standard in the treatment of deep inelastic scattering (DIS) , see e.g. \cite{Thorne:2006qt},
but to our knowledge has never been applied to off-forward processes. 
To the NLO accuracy, we observe the similar simplification as in DIS, allowing one to put the perturbatively generated  
charm quark LCDA to zero at the $c$-quark mass scale, $\mu^2=m_c^2$, which is the basis of the classical ACOT scheme for DIS~\cite{Aivazis:1993pi}.

The presentation is organized as follows. Section 2 summarizes the QCD factorization formalism for transition form factors.
It contains all relevant definitions and notation. The results for the coefficient functions are presented up to two-loop
accuracy, and also the treatment of heavy-quark contributions is explained. Section 3 is devoted to the calculation of the 
scale dependence of the LCDAs to three-loop accuracy. Explicit expressions for the matrix of flavor-singlet anomalous
dimensions are collected in the Appendix.
Section 4 contains our numerical analysis. Predictions are presented for a few typical LCDA models 
under the state mixing assumption, and several scenarios for the
``intrinsic'' gluon LCDA.
Section 5 is reserved for a summary and outlook.

\section{QCD factorization}
Meson transition form factors play a distinguished role in the QCD theory of hard exclusive reactions
\cite{Chernyak:1977as,Chernyak:1977fk,Radyushkin:1977gp,Efremov:1978rn,Efremov:1979qk,Lepage:1979zb,Lepage:1980fj,Brodsky:1980ny}
since the leading contribution at large momentum transfers 
in $\gamma\gamma^\ast\to M$ amplitudes is present at tree level already. 
This is a crucial advantage as compared to, e.g., electromagnetic form factors in which case the leading contribution starts
at order $\alpha_s$ and has to win against nonfactorizable contributions that are power suppressed but do not involve a small 
loop factor $\alpha_s/\pi$. Hence the factorization regime in transition form factors is expected to be achieved much 
earlier, and the leading-twist description has significantly better accuracy compared to other processes. 
It is therefore widely accepted that transition form factors provide one with the best avenue to extract information 
on the hadron wave functions at small transverse distances between the constituents --- the LCDAs.

We define quark and gluon LCDAs as matrix elements of nonlocal light-ray operators
\begin{align}
 & \hspace*{-1cm}\langle 0| \bar q (z_1n)\slashed{n}\gamma_5q (z_2n) |M(p)\rangle
\notag\\ &= i (pn)\int_0^1 du\, e^{-i(z_1u +z_2\bar u)(pn)}\Phi_M^q(u)\,, 
\notag\\
 & \hspace*{-1cm}\langle 0| G_{n\alpha}^a (z_1n) \tilde G_{n\alpha}^a(z_2n) |M(p)\rangle 
\notag\\ &= (pn)^2\int_0^1 du\, e^{-i(z_1u +z_2\bar u)(pn)}\Phi_M^g(u)\,, 
\end{align}
where $M=\eta, \eta'$,  $n_\mu$ is an auxiliary light-like vector, $n^2=0$,  $z_1$ and $z_2$ are (arbitrary) real numbers, 
and we use a notation 
\begin{equation}
 z_{21} = z_2-z_1\,, \qquad z_{21}^u = \bar u z_2 + u z_1\,,\qquad \bar u = 1-u\,.
\end{equation}
The physical meaning of the $u$-variable is the momentum fraction carried by the corresponding constituent (antiquark or gluon). 
Further, $\widetilde G_{\mu\nu}$ is the dual gluon field strength tensor 
$\widetilde G_{\mu\nu} = (1/2)\epsilon_{\mu\nu\alpha\beta}G^{\alpha\beta}$ and $G_{n\xi} = G_{\mu\xi} n^\mu$. 
We use the conventions  $\gamma_5 = i\gamma^0\gamma^1\gamma^2\gamma^3$ and 
$\epsilon_{0123} = 1$, following~\cite{Bjorken:1965zz}.
In our normalization
\begin{align}
\int_0^1 du\,\Phi_M^q (u) = F_M^{q}\,, \qquad q =u,d,s\,,  
\end{align}
where  $ F_M^{q}$ are the axial vector couplings for a given quark flavor
\begin{align}
  \langle 0| \bar q \gamma_\mu\gamma_5 q |M(p)\rangle & = i p_\mu F_M^{q}\,, \qquad q = u,d,s\,.
\label{Fuds}
\end{align} 
The quark LCDAs are symmetric and the gluon LCDA is antisymmetric under the substitution $u\to\bar u$
\begin{align}
 \Phi_M^q(\bar u) = + \Phi_M^{q}(u)\,,  
\qquad
\Phi_M^g(\bar u) = - \Phi_M^{g}(u)\,. 
\end{align}
The transition FFs  $F_M(q^2)$ are real functions in Euclidean region $Q^2 = -q^2 >0$ and can be continued analytically $Q^2 \to -q^2 -i\epsilon$
to  Minkowski region $q^2>0$. To the leading-twist accuracy
\begin{align}
 Q^2 F_M(Q^2) &= \sum\limits_{q} \int_0^1\!du \, \mathbb{T}_q(u,\mu/Q,\alpha_s(\mu)) \Phi^q_M(u,\mu)
\nonumber\\ &\quad +  \int_0^1\!du \, \mathbb{T}_g(u,\mu/Q,\alpha_s(\mu)) \Phi^g_M(u,\mu)\,,  
\end{align}
where $\mu$ is the factorization scale. We tacitly assume using dimensional regularization with minimal subtraction.

The coefficient functions (CFs) can be decomposed in nonsiglet (NS) and pure-singlet (PS) contributions with 
different dependence on electric charges:
\begin{align}
  \mathbb T_q &= e_q^2 T_{NS} + \sum_{q'=u,d,s,c} e^2_{q'} T_{PS}\,, 
\notag\\ \mathbb T_g &= \sum_{q'=u,d,s,c} e^2_{q'} T_g^{(q')} + e_c^2 \delta T_g^{(c)}+ e_b^2T_g^{(b)},    
\label{mathbbCFs}
\end{align}
where $\delta T_g^{(c)}$ is power-suppressed $\mathcal O(m_c^2/Q^2)$
remainder of the $c$-quark contribution after changing to the $n_f=4$ scheme at $\mu_0^2 =m_c^2$ (see below).
We do not see a compelling reason to go over to the $n_f=5$ scheme at the $b$-quark mass scale  
since the $Q^2$ range accessible in form factor measurements is limited 
\footnote{The resonance contribution $e^+e^-\to \Upsilon(4S)\to \eta^{(')}\gamma$ is claimed in Ref.~\cite{BaBar:2006ash} to be very small
due to the estimated tiny branching fraction $\mathcal{B}( \Upsilon(4S)\to \eta^{(')}\gamma)$. Hence we neglect it.}.

The CFs $T_{NS}$ and $ T_{PS} $ \eqref{mathbbCFs} do not depend on the quark flavor and have a regular
perturbative expansion
\begin{align}
   T_{r} =  T_{r}^{(0)} + a_s T_{r}^{(1)} +  a^2_s T_{r}^{(2)} \,, \qquad a_s = \frac{\alpha_s}{4\pi}\,. 
\end{align}
The PS CF first appears at two loops, so that $T_{PS}^{(0)}= T_{PS}^{(1)}=0$. 
Explicit expressions for massless quarks to two-loop accuracy can be found in 
Refs.~\cite{Braun:2020yib,Gao:2021iqq,Braun:2021grd,Braun:2022bpn,Ji:2023xzk}.

The contribution of light quarks $\ell = u,d,s$ to the one-loop gluon CF is well known,
\begin{align}
  T_g^{(\ell)(1)}(u) &=  - \frac{4 \ln u}{\bar u^2}\left(\frac{1}{u}-3 +\frac12 \ln u +L\right) - (u\leftrightarrow \bar u)\,,
\end{align} 
where  $L= \ln Q^2/\mu^2$.
The one-loop heavy-quark contribution $h=c,b$ was calculated in 
\cite{Noritzsch:2003un,Agaev:2014wna}. We have repeated this calculation and confirm the 
result:
\begin{align}
  T_g^{(h)(1)}(u) &= \frac{2}{u\bar u} \biggl\{
\frac{1}{\bar u} \Big[
- u  \mathcal L_h^2(u) + 2  (3u-1)\beta_h(u)\mathcal L_h(u) 
\notag\\&\quad +  \mathcal L_h^2(1)  - 4 \beta_h(1)\mathcal L_h(1) \Big] - (u\leftrightarrow\bar u)
\biggr\},
\label{Th}
\end{align}
where
\begin{align}
\beta_h(u) &= \sqrt{1+\frac{4m_h^2}{u Q^2}}\,, \qquad 
 \mathcal L_h(u) = \ln\left(\frac{\beta_h(u)+1}{\beta_h(u)-1}\right)\,.
\end{align}
The logarithms $\sim a_s(\mu)\ln \mu^2/m_c^2$ can be resummed at the leading power using a method 
known as the variable flavor number scheme (VFNS). Expanding the expression in Eq.~\eqref{Th}
in powers of the quark mass one can write the result as 
\begin{align}
  T_g^{(c)(1)}(u) &=  T_g^{(\ell)(1)}(u) 
 +   \ln\frac{\mu^2}{m_c^2}\Big( \frac{\ln \bar u}{u^2} - \frac{\ln u }{\bar u^2}\Big)
\notag\\ &\quad
 +  \delta T_g^{(c)(1)}(u)\,, 
\end{align} 
where $\delta T_g^{(c)(1)}(u) =\mathcal{O}(m_c^2/Q^2)$ collects all power-suppressed contributions.
The first term on the r.h.s. of this expression is taken into account as the $c$-quark
contribution to the gluon CF (the first term in \eqref{mathbbCFs}) in the approximation that the $c$-quark mass is set 
to zero. The second term can be rewritten using that
\begin{align}
  4 \left[\frac{\ln \bar u}{u^2} - \frac{\ln u }{\bar u^2}\right] =   \int_0^1\!dv\,  T^{(0)}_{NS}(v) H_{qg}(v,u)\,,
\end{align}
where  $T^{(0)}_{NS}(v) = 1/v +1/\bar v$ and
$H_{qg}$ is the $q\to g$ evolution kernel
\begin{align}
H_{qg}(v,u) = -4 \biggl[ \frac{v}{u^2} \Theta(u-v) -  \frac{\bar v}{\bar u^2} \Theta(v-u)\biggr]. 
\end{align}
This contribution can, therefore, be reinterpreted as a contribution of the $c$-quark LCDA 
\begin{align}
\Phi^c_M(v,\mu) &
=   a_s \ln\frac{\mu^2}{m_c^2} \int_0^1\!du \, H_{qg}(v,u)\Phi^g_M(u,\mu)
\label{cLCDA}
\end{align}
generated through the $c$-quark-gluon mixing starting at $\mu_0^2 =m_c^2$. 

Note that in the chosen $\widebar{\text{MS}}$ subtraction scheme at one-loop order 
there is a logarithm $\ln\frac{\mu^2}{m_c^2}$ but no constant term. 
This is the same simplification as found in DIS~\cite{Aivazis:1993pi}, and it allows us to
set the perturbatively generated $c$-quark LCDA to zero at the initial scale $\mu_0^2 =m_c^2$.
\footnote{This simplicity is special for NLO and probably accidental, cf.~\cite{Aivazis:1993pi}.  
 The construction of a consistent VFNS at NNLO is considerably more involved, see, e.g., Ref.~\cite{Thorne:2006qt} for a detailed
 discussion for DIS. Such construction is beyond the scope of our work as the two-loop gluon CF with 
 massive quarks is not known yet.}
  
 The LCDAs can be expanded in  Gegenbauer polynomials so that the expansion coefficients have autonomous evolution at one loop
\begin{flalign}
   \Phi^{q}_M(u,\mu) &=  F^{(q)}_M\,  6 u\bar u \sum\limits_{n=0,2,\ldots} \!a^{q,M}_{n}(\mu) C^{3/2}_n(2u-1),
\notag\\ 
  \Phi^{g}_M(u,\mu) &=  F^{0}_M \, 30 u^2\bar u^2 \sum\limits_{n=2,\ldots}\! b^M_{n}(\mu) C^{5/2}_{n-1}(2u-1).
\label{LCDA}
\end{flalign}
In our normalization $a_0^{q,M} = 1$ for the light quarks, $q=u,d,s$. 
We have chosen to normalize the gluon LCDA to the three-flavor-singlet quark coupling
\begin{align} 
F_M^0 = ( F_M^u +  F_M^d + F_M^{s})/\sqrt{6}\,,
\label{F0M}
\end{align} 
it seems that there exists no established convention.
The (dimensionless) coefficients $a_n$ and $b_n$ for flavor-singlet contributions mix with each other, and beyond leading 
order also with lower coefficients $a_{n'}$ and $b_{n'}$, $n' = 0, \ldots, n-2$.
In what follows we refer to these coefficients as shape parameters. The values of shape parameters at a low scale $\mu_0$
encode all nonperturbative information on the LCDAs. We come back to their discussion in Section~4.


\begin{table*}[t]
\renewcommand{\arraystretch}{1.3}
\begin{center}
\begin{tabular}{|l|l|l|l|l|l|} \hline   
  n & 0 & 2 & 4 & 6 & 8 
\\ \hline\hline
 $ M_{\scriptscriptstyle NS}^{(0)}(n)$ 
& $ 6$
& $ 6$
& $ 6$
& $ 6$
& $ 6$
\\ \hline
 $ M_{\scriptscriptstyle NS}^{(1)}(n)$ 
& $ -72.$
& $ 27.4444 - 33.3333 L$
& $ 91.0844 - 48.5333 L $
& $ 140.942 - 58.6857 L $
& $ 182.65 - 66.3492 L$
\\ \hline
\multirow{2}{*}{$ M_{\scriptscriptstyle NS}^{(2)}(n)$}
& $  - 315.005  + 147.791 L$
& $  - 993.597  - 88.2495 L$
& $  - 781.498  - 664.354 L$
& $  - 121.054  - 1308.01 L$
& $     772.541 - 1855.59 L$
\\
& 
& $ + 109.259 L^2$
& $ + 220.557 L^2 $   
& $ + 316.344 L^2$
& $ + 400.026 L^2$
\\ \hline\hline
 \multirow{2}{*}{$ M_{\scriptscriptstyle PS}^{(2)}(n)$}
 & $ 180.107 - 52.6379 L $
 & $ 43.8174 - 13.5878 L $ 
 & $ 23.5022 - 6.68667 L $
 & $ 15.0973 - 4.00137 L $ 
 & $ 10.6724 - 2.67838 L $
\\
 & 
 & $  + 1.11111 L^2 $ 
 & $  + 0.497778 L^2$
 & $  + 0.27551 L^2$ 
 & $  + 0.173827 L^2$
\\ \hline\hline
 $  M_g^{(1)}(n)$
 & $ - $
 & $ - 76.3889 + 16.6667 L$ 
 & $ - 103.289 + 18.6667 L$
 & $ - 118.952 + 19.2857 L $ 
 & $ - 129.982 + 19.5556 L$
\\ \hline
\multirow{2}{*}{ $ M_g^{(2)}(n)$}
 & $  - $
 & $  -3283.93 + 1608.94 L$ 
 & $  -5909.67 + 2754.59 L$
 & $  -8190. + 3607.69  L $ 
 & $  -10215.4 + 4296.89 L$
\\
& $ $
 & $  -212.963 L^2 $ 
 & $  -323.763 L^2$
 & $  -390.214 L^2$ 
 & $  -437.45  L^2$
\\ \hline
\end{tabular}
\end{center}
\caption[]{\label{table:1}
\sf Gegenbauer moments \eqref{moments} 
of the two-loop coefficient functions for light quarks and gluons, 
$n_f=4$, $L= \ln (Q^2/\mu^2)$}
\renewcommand{\arraystretch}{1.0}
\end{table*}

Thanks to conformal symmetry, the contributions  of each $C^{5/2}_{n-1}$ polynomial in the expansion of the gluon LCDA are mapped 
(at one loop) into the contributions of $C^{3/2}_n$ polynomials  in the  $c$-quark LCDA $\Phi^c_M(u)$:   
\begin{align}
 \Phi_M^{c}(u) &\sim  \int_0^1dv\, H_{qg}(u,v) \Phi^g_M(v) 
\notag\\
 &=  F^{0}_M \, 6 u\bar u \sum\limits_{n=2,4,\ldots}\!\! 
\frac{10 n(n+3)}{3(n\!+\!1)(n\!+\!2)} 
b_n^M C^{3/2}_n(2u-1)\,. 
\end{align}
Note that the $n=0$ term is missing at one loop order, but it will be generated at two loops.
In the asymptotic limit $\mu\to \infty$ assuming that there is no intrinsic charm contribution,
$F^c(\mu_0)=0$, one obtains 
\begin{align}
 F^c_M(\mu\to\infty, n_f=4) &= - \frac{6\sqrt{6} C_F }{\beta_0} a_s(\mu_0) F^0_M(\mu_0) 
\notag\\&\quad
+ \mathcal{O}(a^2_s(\mu_0)). 
\end{align}

We define the Gegenbauer moments of the light quark and gluon CFs as 
\begin{align}
 M_{NS}^{(k)}(n) &= 6 \int_0^1\!du\, T_{NS}^{(k)}(u)\, u \bar u\, C_n^{3/2}(2u-1),
\nonumber\\
 M_{PS}^{(k)}(n) &= 6 \int_0^1\!du\, T_{PS}^{(k)}(u)\, u \bar u\, C_n^{3/2}(2u-1), 
\nonumber\\
 M_{g}^{(k)}(n) &= 30 \int_0^1\!du\, T_{g}^{(\ell)(k)}(u)\, u^2 \bar u^2\, C_{n-1}^{5/2}(2u-1). 
\label{moments}
\end{align} 
The numerical values for $n=0,2,4,6,8$ at tree level, one-loop and two-loop order for 
$n_f=4$ are collected in Table~\ref{table:1}. The complete analytic expressions 
for arbitrary $n_f$ can be found in the ancillary file.
The heavy-quark contributions involving $\delta T_g^{(c)(1)}(u, m_c^2/Q^2)$ and $T_g^{(b)(1)}(u,m_b^2/Q^2)$ 
are calculated from the expressions presented above directly in momentum fraction space. 

Note that the CFs are scheme-dependent beyond leading order. We use the results for 
$T_{g}^{(\ell)(2)}(u)$ and  $T_{PS}^{(2)}(u)$ from Ref.~\cite{Ji:2023xzk} in Larin's scheme,
and $T_{NS}^{(k)}(u)$  from Refs. \cite{Braun:2021grd,Gao:2021iqq}
in the ``physical'' scheme corresponding to anticommuting $\gamma_5$ matrix.
This choice is consistent with the chosen renormalization scheme for the relevant operators, 
see next Section.

\section{Scale dependence of the LCDAs at NNLO}

\subsection{Operator renormalization: general setup}

Our presentation and notations in this section follow closely Ref.~\cite{Braun:2022byg}.
The expressions below are written for the flavor-singlet case.

Let $n^\mu$ and $\bar n^\mu$ be a pair of lightlike vectors, 
\begin{align}
n^2=\bar n^2=0, && (n\cdot\bar n) =1\,,
\end{align}
and $a^\mu$ and $b^\mu$ be mutually orthogonal unit vectors in the transverse plane,
\begin{align}
 (a\cdot n) &= (a\cdot \bar n) = (b\cdot n) = (b\cdot \bar n) = (a\cdot b)=0\,.
\end{align}
In what follows we use the notation $p_+=(p\cdot n)$,  $p_-=(p\cdot  \bar n)$, $p_a = (p\cdot a)$, etc.

Moments of the LCDAs are given by the matrix elements of local operators.
In  noninteger space-time dimensions, $d = 4- 2\epsilon$,  
the leading-twist quark and gluon axial-vector operators can be defined as follows:
\begin{subequations}
\label{OLarin}
 \begin{align}
 \mathcal O^q_n & = i \,\partial_+^{n}\, \sum_{f=1}^{n_f}\bar q^f\, C_{n}^{(3/2)}
\left(\frac{\Dleft-\Dright}{\Dleft+\Dright}\right)\Gamma_+^{ab} q^f,
\\
\mathcal O^g_n  & = 12\,
\partial_+^{n-1}\, G^{a+}\, C_{n-1}^{(5/2)}
\left(\frac{\Dleft-\Dright}{\Dleft+\Dright}\right) G_{b+}\,.
\end{align}
\end{subequations}
Here $D^\mu = \partial^\mu - i g A^\mu$ is the covariant derivative, and
\begin{align}
 \Gamma_+^{ab} = a^\mu b^\nu n^\rho \Gamma_{\mu\nu\rho}\,,
\end{align}
where
\begin{align} 
\Gamma_{\mu\nu\rho}= \frac16(\gamma_\mu\gamma_\nu\gamma_\rho\pm \ldots)
\end{align}
is the antisymmetrized product of three $\gamma$-matrices. 

In four dimensions, $d=4$, the operators $\mathcal{O}_{n}^{q}$,  $\mathcal{O}_{n}^{g}$ can be written 
in terms of the ``standard'' flavor-singlet axial-vector operators
\begin{align}
  \epsilon_{+-ab}\,\mathcal{O}_{n}^{q} &= {O}_{n}^{q}, 
\qquad
  \epsilon_{+-ab}\,  \mathcal{O}_{n}^{g} = {O}_{n}^{g}, 
\end{align}
where
\begin{align}
 {O}_{n}^{q} &=  
\partial^n_+\sum\limits_{f=1}^{n_f} \bar q^f C^{3/2}_n\left(\frac{\Dleft-\Dright}{\Dleft+\Dright}\right)\gamma_+\gamma_5 q^f,
\notag\\
  {O}_{n}^{g} &= 
6 \partial^{n-1}_+ G_{\mu+} C^{5/2}_{n-1}\left(\frac{\Dleft-\Dright}{\Dleft+\Dright}\right) \widetilde{G}^{\mu+}. 
\label{O5}
\end{align}

The quark and gluon operators defined in \eqref{OLarin} have spin $N=n+1$ and mix with each other 
under renormalization for even $n=2,4,6,\ldots$,
\begin{align}
[\mathcal O_n^\alpha] &= Z_n^{\alpha\beta} \mathcal O_n^\beta + \text{total derivatives}\,, 
%
\end{align}
where $\alpha,\beta \in  \{q, g\}$. Here and below $[\ldots]$ stands for renormalization in the $\overline{\text{MS}}$ scheme.
For odd $n=1,3,\ldots$ and for $n=0$ only the quark operator exists and there is no mixing.
In the following we assume that $n$ is even.

Since operators containing total derivatives do not contribute to forward matrix elements,
these matrix elements satisfy the renormalization group equation (RGE) of the form
\begin{align}
\Big(\big(\mu\partial_\mu 
+\beta(a_s)\partial_{a_s}\big) \delta^{\alpha\beta}+\gamma_n^{\alpha\beta}(a_s)\Big) 
\langle p| [\mathcal O^\beta_n]|p\rangle =0\,,
\label{RGE_forward}
\end{align}
where $\beta(a_s)$ is the QCD $\beta$-function
\begin{align}
 \beta(a_s) = -2 a_s (\beta_0 a_s + \beta_1 a_s^2 + \ldots).
\end{align}
The first three coefficients that are relevant for our analysis are~\cite{Tarasov:1980au}
\begin{align}
\beta_0 &
=11-\frac23 n_f\,, 
\notag\\
\beta_1 &
=102-\frac{38}3 n_f\,,
\notag\\
\beta_2&
=\frac{2857}2-\frac{5033}{18}n_f+\frac{325}{54} n_f^2.
\end{align}
The anomalous dimensions (ADs)
\begin{align}\label{AD}
\gamma^{\alpha\beta}_n =-\mu\partial_\mu Z_n^{\alpha\alpha'} (Z_n^{-1})^{\alpha'\beta},
\end{align}
are $2\times 2$ matrices
\begin{align}
 \gamma_n &= 
\begin{pmatrix} \gamma_n^{qq} & \gamma_n^{qg} \\ \gamma^{gq}_n & \gamma^{gg}_n \end{pmatrix}
 = a_s \gamma_n^{(1)} + a_s^2 \gamma_n^{(2)} +\ldots \,
\label{gamma_n}
\end{align}
that are known to three-loop accuracy for all $n$~\cite{Moch:2014sna,Behring:2019tus}.
The ADs depend only on the coupling constant, so that
the RGE \eqref{RGE_forward} in $d=4-2\epsilon$ dimensions has the same form as in $d=4$, 
but with the $d$-dimensional $\beta$-function $\beta(a_s) = -2 a_s (\epsilon + \beta_0 a_s + \ldots)$.

In processes with nonzero momentum transfer, mixing with operators containing total derivatives
has to be taken into account,
\begin{align}\label{Omn}
 \mathcal O^\alpha_{mn} & = \partial_+^{n-m}\mathcal O^\alpha_{m}\,, \qquad m= n-2, n-4, \ldots\,.
\end{align}
Since $[\partial_+^{n-m}\mathcal O^\alpha_{m}] = \partial_+^{n-m}[\mathcal O^\alpha_{m}] $,
we can write
\begin{align}
[\mathcal O_{mn}^\alpha] = \sum_{k=0,2,\ldots,m} Z_{mk}^{\alpha\beta} \mathcal O_{kn}^\beta,
\end{align}
which has the same form for all $n$, so that this subscript is essentially redundant.

We will use matrix notation
\begin{align}
  \vec{\mathcal O}_n &=
\begin{pmatrix}   \mathcal O^q_n \\   \mathcal O^g_n \end{pmatrix} 
\label{Ovector}
\end{align}
and
\begin{align}
\Oboldcal_n =\begin{pmatrix}
\vec{\mathcal O}_{0n}\\
\vec{\mathcal O}_{2n}\\
\vdots
\\
\vec{\mathcal O}_{nn}
\end{pmatrix}\,,\qquad
%
\Zboldcal_n =\begin{pmatrix}
Z_{00} & 0 &\cdots &0 &\\
Z_{20} & Z_{22} & \cdots &0\\
\vdots & \vdots &\ddots &\vdots\\
Z_{n0} & Z_{n2} &\cdots& Z_{nn}
\label{Oboldcal}
\end{pmatrix},
\end{align}
where each entry $Z_{mk}$ is a $2\times 2$ matrix $Z_{mk}^{\alpha\beta}$
except for $Z_{00}$ which is a number, and $Z_{m0}$, $m>0$, which are vectors.
Note that the matrix $\Zboldcal_m$ for $m<n$ is a principal submatrix of $\Zboldcal_n$:  the subscript
only specifies the size of the matrix while the entries do not depend on it.
In this notation, the RGE for $[\Oboldcal_n]=\Zboldcal_n \Oboldcal_n$ takes the form
\begin{align}
\big(\mu\partial_\mu+\beta(a_s)\partial_{a_s} + \boldsymbol{\gamma}_n(a_s)\big) [\Oboldcal_n]=0\,,
\end{align}
where
\begin{align}\label{boldgamma}
%
\bm{\gamma}_n(a_s) =\begin{pmatrix}
\gamma_{00} & 0 &\cdots &0 &\\
\gamma_{20} & \gamma_{22} & \cdots &0\\
\vdots & \vdots &\ddots &\vdots\\
\gamma_{n0} & \gamma_{n2} &\cdots& \gamma_{nn}
\end{pmatrix}.
\end{align}
This matrix has the same structure as \eqref{Oboldcal}:
each entry $\gamma_{mk}$ is a $2\times 2$ matrix $\gamma_{mk}^{\alpha\beta}$, $\alpha,\beta = q,g$,
except for $\gamma_{00} = \gamma_{00}^q$ which is a number and  $\gamma_{m0}$ which are $2\times 1$ vectors. 
The diagonal blocks $\gamma_{nn}$ in this matrix are the usual  ADs~\eqref{gamma_n}, 
$\gamma_{nn}\equiv\gamma_n$. The off-diagonal blocks $\gamma_{km}$, $k>m$ 
describe  mixing with the total derivatives operators \eqref{Omn} and,
in the basis~\eqref{OLarin},  start at order $O(a_s^2)$:
\begin{align}\label{gamma-off}
 \gamma_{km}(a_s)& =a_s^2\gamma_{km}^{(2)}+ a_s^3\gamma_{km}^{(3)}+\ldots\,, \qquad k>m\,.
\end{align}

The renormalization scheme described above 
is  an extension of the well-known Larin's prescription \cite{Larin:1993tq}.         
This scheme is self-consistent in QCD. For the flavor-nonsinglet operators, however, it is customary to 
apply an additional finite renormalization to convert the results from  Larin's scheme 
to the scheme with an anticommuting $\gamma_5$ matrix such that the ADs for vector and axial-vector operators coincide to all orders in perturbation theory. 
We are not aware of any compelling prescription for doing such a finite renormalization for the flavor-singlet case.  
Nevertheless, for consistency with the treatment of flavor-nonsinglet contributions, we will apply the same 
transformation for flavor-singlet quark operators as well:
\begin{align}
  [\mathbf {O}_{mn}^{q}] & =  \sum\limits_{k=2,4,\ldots,m} \mathrm U_{mk}  [\mathcal{O}_{kn}^{q}]\,.  
\end{align}
In matrix notation \eqref{Ovector}
\begin{align}
   [\vec{\mathbf O}] & = \widehat{\mathrm U}\,[\vec{\mathcal{O}}]\,,
\qquad
\widehat{\mathrm U} = 
\begin{pmatrix}
 \mathrm U & 0 \\ 0 & 1
\end{pmatrix}.
\end{align}
The RGE for ``rotated'' operators takes the form
\begin{align}
\big(\mu\partial_\mu+\beta(a_s)\partial_{a_s} + \widetilde{\boldsymbol{\gamma}}\big) [\vec{\mathbf O}]=0\,,
\end{align}
where the AD matrix $\widetilde{\boldsymbol{\gamma}}$ is related to the one in Larin's scheme, 
${\boldsymbol{\gamma}}$, as
\begin{align}
 \widetilde{\boldsymbol{\gamma}} &= \widehat{\mathrm U} \boldsymbol{\gamma} \widehat{\mathrm U}^{-1}
- \beta(a_s) \partial_{a_s} \widehat{\mathrm U} \,\widehat{\mathrm U}^{-1}. 
\end{align}
The matching kernel $\mathrm{U}$ for the forward matrix elements (of flavor-nonsinglet operators) 
is available from \cite{Moch:2014sna}
and for the off-forward matrix elements from \cite{Braun:2021tzi}. 
It can conveniently be written as~\cite{Braun:2021tzi}
\begin{align}
\mathrm U =\exp\left\{\!-\frac{a_s}{2\beta_0} V_1-\frac{a_s^2}{4\beta_0}\left( V_2
-\frac{\beta_1}{\beta_0} V_1\right)\right\}\! +\mathcal{O}(a_s^3)
\end{align}
such that 
\begin{align}
 \widetilde{\boldsymbol{\gamma}} &= \widehat{\mathrm U} \boldsymbol{\gamma} \widehat{\mathrm U}^{-1}
-\begin{pmatrix}
 V(a_s) & 0 \\ 0 & 0
\end{pmatrix},
\end{align}
where $V(a_s)= a_s^2 V_1+ a_s^3 V_2+\mathcal{O}(a_s^4)$. The matrix $V_1$ is diagonal
\begin{align}
[V_1]_{nm}=\delta_{nm}\frac{16 C_F \beta_0}{(n+1)(n+2)}\,,
\end{align}
whereas $V_2$ is more complicated and also contains an off-diagonal part, 
see Ref.~\cite{Braun:2021tzi} for the explicit expressions.
Using this representation we obtain
\begin{align}\label{tildegamma}
 \widetilde{\boldsymbol{\gamma}}^{(1)} &= {\boldsymbol{\gamma}}^{(1)},
 \notag\\
  \widetilde{\boldsymbol{\gamma}}^{(2)} &= {\boldsymbol{\gamma}}^{(2)} +\frac1{2\beta_0}\left[ {\boldsymbol{\gamma}}^{(1)}, V_1\right] - V_1,
  \notag\\
  \widetilde{\boldsymbol{\gamma}}^{(3)} &= {\boldsymbol{\gamma}}^{(3)} +\frac1{2\beta_0}\left[ {\boldsymbol{\gamma}}^{(2)}, V_1\right] 
  +\frac1{4\beta_0} \left[ {\boldsymbol{\gamma}}^{(1)}, V_2-\frac{\beta_1}{\beta_0} V_1\right]
  \notag\\
  &\quad +\frac1{8\beta_0^2}\left[ \left[ {\boldsymbol{\gamma}}^{(1)}, V_1\right],V_1\right] -V_2\,,
\end{align}
where $[\ldots,\ldots]$ stands for the commutator.
The two-loop off-diagonal entries \eqref{gamma-off} of the AD matrix for the operators in 
question, $\gamma_{km}^{(2)}$, have been obtained  in Ref.~\cite{Belitsky:1998gc}. 
We have calculated the three-loop ADs $\gamma_{km}^{(3)}$ which are our main new input.
 
\subsection{The three-loop AD matrix}

We use the technique developed in Ref.~\cite{Braun:2022byg}. 
The starting point is that conformal symmetry of QCD at quantum level is restored at the
Wilson-Fisher critical point \cite{Wilson:1973jj} at noninteger space-time dimension $d= 4-2\epsilon_\ast$,  
\begin{align}
   \epsilon_\ast(a_s) = -\beta_0 a_s -\beta_1 a_s^2 - \ldots \,.
\label{crit}
\end{align}
At the critical point, the two-point correlation functions 
of multiplicatively renormalizable operators with different 
ADs vanish to all orders of perturbation theory~\cite{Polyakov:1970xd}
\begin{align}
      \langle [\mathcal{O}]_n (x) [\mathcal{O}]_m (0) \rangle \sim \delta_{nm}\,, \qquad x \slashed{=} 0\,, 
\label{method}
\end{align} 
where $\langle\ldots\rangle$ stands for the vacuum expectation value.
As shown in Ref.~\cite{Braun:2022byg}, this condition allows one to find the eigenvectors of the 
renormalization group (RG) equation in the chosen operator basis from a calculation of the 
corresponding unrenormalized correlation functions with $m \le n$.
Since the eigenvalues (ADs) are known~\cite{Moch:2014sna,Behring:2019tus}, 
this information is sufficient to restore the complete mixing matrix.
Importantly, the ADs of composite operators in minimal subtraction schemes do 
not depend on $\epsilon$ by construction and are the same for the physical $d=4$ and 
the critical $d=4-2\epsilon_\ast$ space-time dimensions. Thus the calculated mixing matrix for the 
leading-twist operators at the critical point coincides identically with that in the physical theory 
in four dimensions~\cite{Braun:2013tva,Braun:2016qlg,Braun:2017cih,Ji:2023eni}. 

We have calculated the correlation functions \eqref{method} 
for the operators \eqref{OLarin} with  $n,m = 2,4,6,8$ to three-loop accuracy in $4-2\epsilon$ dimensions
for a generic  gauge group. All the diagrams were generated with the help  of
QGRAF \cite{QGRAF} and evaluated with  FORM \cite{Vermaseren:2000nd}
programs MINCER \cite{Larin:1991fz}  and  COLOR \cite{COLOR}.
Using these expressions we determined the off-diagonal part of the AD (mixing) matrix
for C-parity even flavor-singlet axial-vector operators.
The present application differs from Ref.~\cite{Braun:2022byg} by a Dirac/Lorentz structure of the 
operators only. This difference makes the coding somewhat more involved and the calculation requires more 
computer time but otherwise the procedure remains the same, so that we omit the details.
   
Explicit expressions for the three-loop AD matrices for QCD,  $N_c=3$,
are rather lengthy. They are  collected in the Appendix and also in the ancillary file in Mathematica format.

\subsection{NNLO RG equation}

For the discussion of the scale dependence, it is convenient to combine the shape parameters 
of the LCDAs \eqref{LCDA} with the normalization constants. Let
\begin{align}
 A^q_n &= \frac{3(n+1)(n+2)}{2(2n+3)} F^{q} a^{q}_n\,,
\notag\\
B_n  &=  \frac{5n(n+1)(n+2)(n+3)}{4(2n+3)} F^{0} b_n\,,
\end{align}
where we suppress the meson dependence $M= \eta,\eta'$.

The (scale-dependent) parameters  $A^q_n$, $B_n$ are given by the reduced matrix elements 
of local operators \eqref{O5},
\begin{align}
\langle 0| O_n^q |M(p)\rangle &=(i p_+)^{n+1} A_n^q,
\notag\\
\langle 0| O_n^g|M(p)\rangle &=(i p_+)^{n+1}   B_n\,. 
\end{align}  
Also
\begin{align}
 A_n^q &= \int_0^1\! du\,  C^{3/2}_n(2u-1) \Phi_M^q(u)\,,
 \notag\\
 B_n& = 6 \int_0^1\! du\, C^{5/2}_{n-1}(2u-1) \Phi_M^g(u)\,.
\end{align}
We assume $n_f=4$ active flavors, and further decompose the quark matrix elements $A_n^q $ into 
the flavor-singlet and flavor-nonsinglet parts
\begin{align}
 A_n^q &= A_n^{q,\NS} + A_n^{\sing}
\end{align}
with 
\begin{align}
 A_n^{\sing} &= \frac14 \sum\limits_{q=u,d,s,c} A_n^q.
\end{align}
The nonsinglet coefficients $A_n^{q,\NS}$ satisfy the RGE (the same for all flavors)
\begin{align}
\Big(\big(\mu\partial_\mu+\beta(a_s)\partial_{a_s}\big)\delta_{nm} + \boldsymbol{\gamma}^{\NS}_{nm}\Big) 
A_m^{q,\NS} =0\,,
\end{align}
where $\boldsymbol{\gamma}^{\NS}$ is the flavor-nonsinglet AD matrix defined as in Eq.~\eqref{boldgamma}
but with all entries that are numbers rather than $2\times2$ matrices.

It is convenient to separate the contribution of one-loop ADs from the higher orders
\begin{align}
 \boldsymbol{\gamma}^{\NS} =  \boldsymbol{\gamma}^{(1)\NS} + \boldsymbol{\gamma}^{(23)\NS}  
\,,
\end{align}
where $\boldsymbol{\gamma}^{(1)\NS}_{nm} = a_s {\gamma}^{(1)\NS}_{nm} \sim \delta_{nm} $  
is a diagonal matrix and
${\boldsymbol{\gamma}}^{(23)\NS}_{nm}  = a_s^2{\gamma}^{(2)\NS}_{nm}  + a_s^3{\gamma}^{(3)\NS}_{nm}$. 
Using this decomposition, we  write the solution in the form
\begin{align}
A^{\NS}(\mu) = \mathbf{U}(a_s(\mu), a_s(\mu_0)) {\mathbf{V}} (a_s(\mu), a_s(\mu_0)) A^{\NS}(\mu_0),  
\end{align}
where $\mathbf{U}(a,b)$ is a diagonal matrix  $\mathbf{U}_{nm} = \mathbf{U}_{n}\delta_{nm}$ that takes into 
account the one-loop evolution
\begin{align}
\Big(\beta(a)\partial_{a}+ \boldsymbol{\gamma}^{(1)\NS} \Big)\,\mathbf{U}(a,b)=0\,, 
\qquad \mathbf{U}(b,b) = \II\,,
\end{align}
and ${\mathbf{V}}(a,b)$ is the solution to the RGE 
\begin{align}
\Big(\beta(a)\partial_{a} 
+ \mathbf{U}(b,a){\boldsymbol{\gamma}}^{(23)\NS} (a)  \mathbf{U}(a,b) 
\Big)\,{\mathbf{V}}(a,b)=0
\end{align}
with the boundary condition $\mathbf{V}(b,b) = \II$. 

One obtains
\begin{widetext}
\begin{align}
\mathbf{U}_{nm}(a,b)&=\delta_{nm}\exp\left\{-\int_b^a \frac{ds s}{\beta(s)}\gamma_n^{(1)}\right\}
=
\delta_{nm}\left(\frac{a}{b}\right)^{ \frac{\gamma_n^{(1)}}{2\beta_0}} 
\exp\biggl\{ -\frac{(a-b)\beta_1}{2\beta_0^2}
\gamma_n^{(1)} \left( 1+\frac{a+b}{2\beta_0}\left(\frac{\beta_2}{\beta_1}-\frac{\beta_1}{\beta_0}\right)\right)
+\ldots\biggr\},
\notag\\
\mathbf{V}_{nm}(a,b) &= \delta_{nm} + b {\gamma}^{(2)}_{nm}  E^{(1)}_{nm}
+ b^2\left({\gamma}^{(3)}_{nm}-{\gamma}^{(2)}_{nm} \frac{\beta_1}{\beta_0}\right)  E^{(2)}_{nm}
+\frac{b^2\beta_1}{2\beta_0^2} {\gamma}^{(2)}_{nm}\left({\gamma}^{(1)}_m - {\gamma}^{(1)}_n  \right)
\left[   E^{(1)}_{nm} -  E^{(2)}_{nm} \right]
\notag\\&\quad
 +b^2\sum_{k}\frac{{\gamma}^{(2)}_{nk} {\gamma}^{(2)}_{km} }{{\gamma}^{(1)}_{m}
 -{\gamma}^{(1)}_{k} +2\beta_0}
            \left[ E^{(2)}_{nm}-  E^{(1)}_{nk}\right]\,,
\end{align}
\end{widetext}
where 
\begin{align}\label{EKNM}
E^{(k)}_{nm} =  \frac{\left(  \frac a b\right)^{\frac{\gamma^{(1)}_m-\gamma^{(1)}_n}{2\beta_0} +k}-1}{\gamma_m^{(1)}-\gamma_n^{(1)}+2k\beta_0}\,.
\end{align}
and we dropped the superscript  ${}^{\NS}$  in order not to overload the notation.

For the flavor-singlet contribution, the one-loop anomalous dimension matrix $\boldsymbol{\gamma}^{(1)\sing}$
is $2\times 2$ block-diagonal. Let $\Rboldcal$ be a matrix that brings $\boldsymbol{\gamma}^{(1)\sing}$
to a diagonal form 
\begin{align}
 \boldsymbol{\gamma}^{\sing} = \Rboldcal\, \widehat{\boldsymbol{\gamma}}^{\sing}\, \Rboldcal^{-1},
\qquad \widehat{\boldsymbol{\gamma}}^{(1)}_{nm} = \delta_{nm} \gamma^{(1)\sing}_n\,.
\end{align}
The solution of the RGE  reads in this case
\begin{align}
\Aboldcal(\mu) &=  \Rboldcal\, \widehat{\mathbf{U}}(a_s(\mu), a_s(\mu_0)) \widehat{\mathbf{V}} 
(a_s(\mu), a_s(\mu_0)) 
\notag\\&\quad
\times  \Rboldcal^{-1}\Aboldcal(\mu_0),  
\end{align}
where $\Aboldcal$ is the vector of the flavor-singlet quark and gluon matrix elements 
$A^{\sing}_n$, $B_n$ arranged
as in Eqs.~\eqref{Ovector}, \eqref{Oboldcal},
and the matrices  $\widehat{\mathbf{U}}$,  $\widehat{\mathbf{V}}$ are given by the same 
expressions as above, with the substitution of the anomalous
dimensions $\bm{\gamma}^{(k),\NS} \mapsto \widehat{\bm{\gamma}}^{(k),\sing}$.  

In this work we take into account contributions of operators with $n = 0,2,4,6,8$. 
In this case one ends up with $5\times5$ and $9\times9$ matrices for the renormalization of flavor-nonsinglet
and flavor-singlet contributions, respectively.

%
\section{Numerical analysis}
%

\subsection{Axial couplings  and models of the LCDAs} 

Low-energy processes involving $\eta$ and $\eta'$ mesons are usually described invoking a certain mixing 
scheme that considers physical mesons as superposition of some fundamental, e.g. flavor-octet and 
flavor-singlet components. This approach is phenomenologically very successful, but
relying on a particular mixing scheme for the coupling constants is in fact not necessary as they 
can be calculated directly on the lattice with competitive and eventually better precision.
We will use the (updated) RQCD21 results \cite{Bali:2021qem}
\begin{align}
 F_\eta^{u(d)}  (\mu_0)&=  72.43\, \big({}^{2.04}_{1.99}\big)\,\text{MeV}\,,
\notag\\
  F_\eta^s   (\mu_0)  &= -109.12\, \big({}^{4.65}_{5.02}\big)\,\text{MeV}\,,
\notag\\
  F_{\eta'}^{u(d)} (\mu_0) &=  55.48\, \big({}^{3.48}_{2.51}\big)\,\text{MeV}\,,
  \notag\\
  F_{\eta'}^s  (\mu_0) &=  143.22 \, \big({}^{4.57}_{7.65}\big)\,\text{MeV}\,,
\end{align}
where $\mu_0^2 = 2\,\text{GeV}^2$. 
The corresponding values for the three-flavor-singlet couplings \eqref{F0M} are
\begin{align}
 F_\eta^0  (\mu_0) &=  14.76\, \big({}^{3.34}_{3.57}\big)\,\text{MeV}\,,
\notag\\
 F_{\eta'}^0  (\mu_0)&=  103.8 \, \big({}^{2.42}_{3.56}\big)\,\text{MeV}\,.
\end{align}
These numbers are obtained from the ones given in Table 25 in  \cite{Bali:2021qem},
changing the scale setting parameter $1/\sqrt{8 t_0^{ph}} = 475(6)\, \text{MeV}$
to $481.3(2.7)\, \text{MeV}$ \cite{RQCD:2022xux} 
and rescaling the results from $\mu^2 =4\,\text{GeV}^2$ to  $\mu_0^2 =2\,\text{GeV}^2$
using the four-loop QCD $\beta$-function and the three-loop anomalous dimension of 
the flavor-singlet axial current~\cite{Larin:1993tq}.

Since the information on the shape of quark LCDAs of $\eta$, $\eta'$ is barely available, it makes 
sense to start with a ``minimal model'' corresponding to the 
popular Feldmann--Kroll--Stech (FKS) mixing scheme~\cite{Feldmann:1998vh}.
It is motivated by the old observation 
that the vector mesons $\omega$ and $\phi$ are to a very good approximation pure 
$\bar u u + \bar d d$ and $\bar s s$ states and the same pattern is observed in tensor mesons. 
The smallness of mixing between strange and non-strange states 
is a manifestation of the celebrated OZI rule that is phenomenologically very successful. 
If the axial $U(1)$ anomaly is the \emph{only} effect that makes the situation in 
pseudoscalar channels different, it is natural to assume that physical states are related to the 
flavor states by an orthogonal transformation involving a single mixing angle.
The FKS scheme is consistent with the low energy phenomenology, see \cite{Gan:2020aco} for a review, 
and is also supported by the direct lattice calculations of the $\eta,\eta'$ couplings at low scales
\cite{RQCD:2022xux,Ottnad:2025zxq}.

Extending the state-mixing  description based on the FKS scheme from the couplings to the LCDAs is a very strong assumption.
If accepted, it implies that the $\eta$ and $\eta'$ LCDAs have the same shape (at low scales): 
\begin{align}
& a_n^{u(d),\eta}(\mu_0) =  a_n^{u(d),\eta'}(\mu_0)   = a_n^{\ell}(\mu_0)\,,   
\notag\\
& a_n^{s,\eta}(\mu_0) =  a_n^{s,\eta'}(\mu_0)   = a_n^{s}(\mu_0)\,.   
\label{eq:QFmodel1}
\end{align}  
If the FKS mixing scheme for the LCDAs is accepted at a low scale, it will be violated only mildly at higher scales by the 
renormalization group effects (that also generate nonvanishing OZI-violating contributions). 
Figure~1 in Ref.~\cite{Agaev:2014wna} shows a comparison of the space-like $\gamma^*\gamma\to \pi^0$ 
experimental data with the non-strange $\gamma^*\gamma \to \eta_\ell$ form factor
 extracted from the combination of BaBar and CLEO measurements of 
$\gamma^*\gamma\to \eta$ and  $\gamma^*\gamma\to\eta'$ assuming the FKS mixing scheme. Were this scheme exact, the two form factors would coincide in the whole $Q^2$ range, up to tiny isospin breaking corrections.
It is seen that the existing measurements do not contradict the FKS approximation at 
low-to-moderate $Q^2 \lesssim 10$~GeV$^2$, whereas at larger photon virtualities 
the comparison is not conclusive as there are significant discrepancies 
between the BaBar \cite{BaBar:2009rrj} and Belle \cite{Belle:2012wwz} pion data.
The BaBar data taken alone show a dramatic difference between the $\gamma^*\gamma\to\pi^0$ and $\gamma^*\gamma\to \eta_\ell$
form factors at large virtualities which cannot be explained by perturbative evolution effects. If confirmed, 
this difference would be a stark indication that the concept of state mixing is not applicable to the 
$\eta$ and $\eta'$ LCDAs so that the 
equality of higher-order Gegenbauer coefficients \eqref{eq:QFmodel1} is strongly violated already at low scales. 
Settling this discrepancy is urgently needed and is an important physics goal for  
Belle II~\cite{Belle-II:2018jsg}. 

From lattice calculations~\cite{RQCD:2019osh}
\begin{align}
 a_2^{\ell}(\mu_0) &=  a_2^\pi(\mu_0) =  0.135 \big({}^{22}_{23}\big)\,, 
\notag\\
 a_2^{s}(\mu_0)   &=  0.115  \big({}^{24}_{26}\big)\,, 
\label{a2:RQCD}
\end{align}
where the second number is obtained from the result for $a_2^8 \equiv a_2^{\eta_8}$ assuming validity of the FKS
mixing scheme with the same mixing angle as for the coupling constants:
\begin{align}
 \frac{(a_2^s)^2-(a_2^8)^2}{(a_2^8)^2-(a_2^\ell)^2} = 
\frac12 \frac{1}{2 F^2_{K^+}/F^2_{\pi^+}-1} \simeq 0.26\,.
\label{a2FKS}
\end{align}

\begin{table}[t]
\renewcommand{\arraystretch}{1.3}
\begin{center}
\begin{tabular}{c|l|l|l|l|c} \hline   
 Model  & $a_2^{\ell(s)}$ & $a_4^{\ell(s)}$ & $a_6^{\ell(s)}$ & $a_8^{\ell(s)}$ & Reference 
\\ \hline\hline
  \multirow{2}{*}{I} 
& $ 0.134$
& $ 0$
& $ 0$
& $ 0$
& \multirow{2}{*}{\cite{RQCD:2019osh}} 
\\
& $ 0.115$
& $ 0$
& $ 0$
& $ 0$
\\ \hline
 \multirow{2}{*}{II} 
& 0.13435
& 0.05111
& 0.02666
& 0.01629
& $p=0.532$
\\
& 0.11510
& 0.04135
& 0.02079
& 0.01236
& $p=0.588$
\\ \hline
 \multirow{2}{*}{III} 
&0.07895
&0.07433
&0.07005 
&0.02219
& \multirow{2}{*}{\cite{Agaev:2012tm}} 
\\
&0.06642
&0.06337
&0.05961
&0.01892
\\ \hline
 \multirow{2}{*}{IV}
&0.22445
&0.10340
&0.07065
&0.00046
& \multirow{2}{*}{\cite{Cloet:2024vbv}} 
\\
& 0
& 0
& 0
& 0
\\ \hline 
\end{tabular}
\end{center}
\caption[]{
\label{table:2}
\sf Gegenbauer coefficients (shape parameters) for the four considered models of the quark LCDAs at
the scale $\mu_0^2 = 2\,\text{GeV}^2$. For each model, the upper numbers correspond to $a^\ell_n$ and the lower ones to
 $a^s_n$, respectively.}
\renewcommand{\arraystretch}{1.0}
\end{table}

For higher moments, several exploratory lattice calculations 
exist \cite{Cloet:2024vbv,Detmold:2025lyb} but are not yet precise enough.

\begin{figure}[t]
\includegraphics[width=0.49\textwidth]{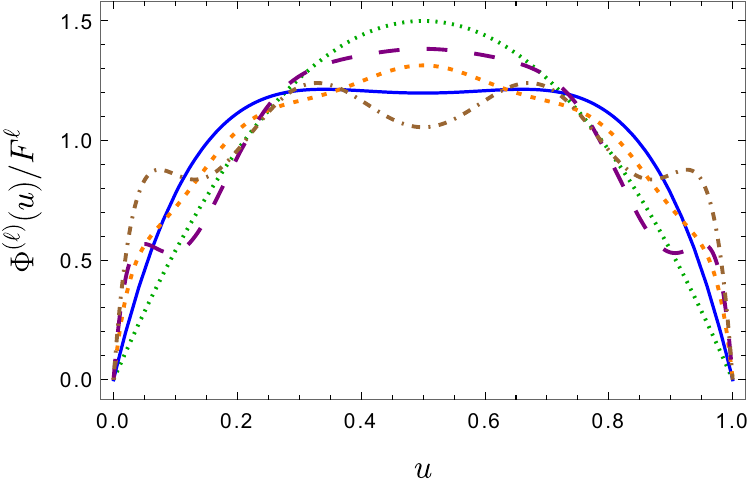}
\caption{
Model shapes of the $u,d$-quark LCDAs at the scale $\mu_0^2 = 2\,\text{GeV}^2$ used in the numerical analysis. 
Model I: solid curve (blue);
Model II: short dashes (orange);
Model III: long dashes (purple);
Model IV: dash-dotted curve (brown).
The asymptotic LCDA $\Phi(u) \sim 6 u(1-u)$ is shown by dots (green) for comparison.
}
\label{figure:LCDAs}
\end{figure}

Numerical results presented below are obtained using four 
phenomenologically acceptable models of the LCDAs, see Table~\ref{table:2} and Fig.~\ref{figure:LCDAs}.
The first model corresponds to taking into account contributions of the Gegenbauer polynomials 
of second order only, with parameters specified in \eqref{a2:RQCD}.
The second model is an extension of the first one, adding a few higher-order polynomials
corresponding to  a power-law ansatz truncated at $n=8$:
\begin{align}
 \Phi^{q}_M = F_M^{q} \frac{\Gamma(2 + 2 p)}{\Gamma(1 + p)^2} u^p (1 - u)^p,
\label{powerLCDA}
\end{align} 

\noindent
where the power $p$ is adjusted to reproduce lattice results~\cite{RQCD:2019osh} for the second moment.
The third model corresponds to the fit of the Belle data \cite{Belle:2012wwz} 
for the space-like $\gamma^\ast\gamma \to \pi$ form factor 
using NLO perturbation theory and light-cone sum rules 
to take into account power-suppressed corrections \cite{Agaev:2012tm}.
For the strange quark contribution in this case we assume the same suppression of 
higher Gegenbauer polynomials as in models I and II. The last model IV corresponds to 
the set of parameters suggested on the basis of the LAMET-type calculation in Ref.~\cite{Cloet:2024vbv}.
An interesting feature of this model is a strong $SU(3)$-flavor violation: Using 
the quoted results for the K-meson~\cite{Cloet:2024vbv}, the leading-order ChPT relation
$a_n^8 = (4 a_n^K-a_n^\pi)/3$ \cite{Chen:2003fp} and Eq.~\eqref{a2FKS} one obtains all 
$a^s_n$ parameters comparable with zero within errors.

For the gluon LCDA there is practically no information available, apart from order-of-magnitude estimates 
\cite{Kroll:2002nt,Kroll:2012gsh}. We will assume 
\begin{align}
 b_2^\eta(\mu_0) =  b_2^{\eta'}(\mu_0) = b_2(\mu_0)\,    
\label{eq:QFmodel2}
\end{align}
with $b_2(\mu_0)$ in the interval $[-0.2, +0.2]$ and put all higher-order $n=4,6,\ldots$ Gegenbauer coefficients to zero (at the initial scale).

\subsection{Form factors}

\begin{table*}[t]
\renewcommand{\arraystretch}{1.2}
\centering
\begin{tabular}{|c|l|l|l|l|l|l|l|}\hline
 flavor                 & scale      &  $\chi_0^{(\alpha)}$     &  $\chi_2^{(\alpha)}$     &  $\chi_4^{(\alpha)}$      &  $\chi_6^{(\alpha)}$        &  $\chi_8^{(\alpha)}$     \\ \hline \hline
\multirow{2}{*}{$u$}    & space-like & 2.286               &  1.678              &  1.464               &   1.374                &  1.340                \\         
                        & time-like  & 2.204 + 0.104 i     & 1.453 + 0.432 i     &  1.321 + 0.528  i    &   1.255 + 0.564 i     &  1.220 + 0.586 i       
\\  \hline\hline 
\multirow{2}{*}{$d,s$}    & space-like & 0.561             & 0.487               & 0.404                & 0.366                  & 0.348                      \\
                        & time-like  & 0.480 + 0.134 i     & 0.303 + 0.210 i     & 0.328 + 0.218 i      & 0.330 + 0.194 i        &  0.323 + 0.170 i  
\\  \hline\hline 
\multirow{2}{*}{$g$}    & space-like &            &  2.143                & 2.304                & 2.283                  &  2.295                                 \\
                        & time-like  &            & -2.451 + 2.643 i      & -0.717 + 5.264 i      & 1.365 + 5.685 i       &  3.322 + 4.059 i  
\\  \hline\hline  
\end{tabular}
\caption{\label{table:3}
Coefficients (\ref{generic}) of the contributions of different Gegenbauer polynomials in the expansion of DAs to the 
transition form factors at the time-like $Q^2 = -s = -112$~GeV$^2$. 
The corresponding space-like coefficients for $Q^2 = 112$~GeV$^2$ are also given for comparison. All numbers are dimensionless.}
\renewcommand{\arraystretch}{1.0}
\end{table*}

We use the RunDec Mathematica package \cite{Chetyrkin:2000yt} for the running strong 
coupling. The heavy quark masses are chosen as $m_c= 1.41$~GeV, $m_b=4.8$~GeV for charm and bottom, respectively.
As already mentioned above, we use $\mu_0^2 = m_c^2 = 2\,\text{GeV}^2$ 
and put charm quark LCDA to zero at this scale. The form factors are evaluated for the factorization scale
 $\mu^2=|Q^2|$ as the central value, and we vary it by a factor two to test the scale dependence.

The transition form factors are linear functions of the parameters 
of the LCDAs \eqref{LCDA} at the initial scale so that the results can be written as 
\begin{align}
Q^2 F_M(Q^2) &= F_M^{u}(\mu_0) \Big[\chi^{(u)}_0 + \sum\limits_{n=2,4,\ldots} \chi^{(u)}_n a_{n}^{(u),M}(\mu_0)\Big] 
\notag\\&\quad
 + F_M^{d}(\mu_0) \Big[\chi^{(d)}_0 + \sum\limits_{n=2,4,\ldots} \chi^{(d)}_n a_{n}^{(d),M}(\mu_0)\Big] 
\notag\\&\quad
 + F_M^{s}(\mu_0) \Big[\chi^{(s)}_0 + \sum\limits_{n=2,4,\ldots} \chi^{(s)}_n a_{n}^{(s),M}(\mu_0)\Big]
\notag\\&\quad
 + F_M^{0}(\mu_0)  \sum\limits_{n=2,4,\ldots} \chi^{(g)}_n b_{n}^{M}(\mu_0)\,,
\label{generic}
\end{align}
where the axial couplings $F_M^{(q)}, F_M^{(0)}$ are defined in \eqref{Fuds},\eqref{F0M}, $a_n$ and $b_n$ are the 
coefficients in the Gegenbauer expansion \eqref{LCDA} of the quark and gluon LCDAs at the scale $\mu_0$,
and $\chi^{(q)}_n,\chi^{(g)}_n$ are dimensionless numbers that depend on $Q^2$ and involve full complexity
of the two-loop CFs and the three-loop evolution. 
Importanty, these coefficients do not depend on the meson, $M = \eta$ or $M =\eta'$. 
Also $\chi^{(d)}_n = \chi^{(s)}_n$ but $\chi^{(u)}_n\slashed{=}\chi^{d(s)}_n$. 
In the isospin symmetry limit $F_M^{(u)} = F_M^{(d)}$ and $a_{n}^{(u),M}=  a_{n}^{(d),M}$.

The numerical values of the coefficients $\chi^{(\alpha)}_n$,  $n\le 8$, for the time-like $Q^2 = - 112\,\text{GeV}^2$ and the factorization scale $\mu^2 = |Q^2|$
are collected in Table~\ref{table:3}. The corresponding space-like coefficients for $Q^2 = 112$~GeV$^2$ are also given for comparison.
This form of presentation of the results is very convenient as it does not rely on any LCDA model.
In particular the expansion in \eqref{generic} does not rely on the state mixing assumption.
In the FKS mixing scheme used in this work $a_{n}^{(q),\eta}(\mu_0)= a_{n}^{(q),\eta'}(\mu_0)$ and  $b_{n}^{\eta}(\mu_0)= b_{n}^{\eta'}(\mu_0)$,
but this assumption can be relaxed.  

The difference between the NNLO and the NLO results is small, of the order of $1-1.5\%$, and
the uncertainty of the NNLO predictions for absolute values of the form factors due to the variation 
of the factorization scale in the interval $|Q^2|/2 < \mu^2 < 2|Q^2|$ is less than 1\% in all cases with an exception
for $\gamma^\ast\to \eta'\gamma$  with a large negative gluon LCDA, $b_2(\mu_0)=-0.2$, in which case
the scale uncertainty is around 2\%. This means that the perturbative expansion is converging rapidly and
the achieved accuracy is sufficient for a fully quantitative description. 

The size of the $c$-quark contribution (terms involving $c$-quark electric charge $e_c^2$) depends strongly
on the assumed gluon LCDA and is in general much larger for $\gamma^\ast\to \eta'\gamma$, as
can be expected. For the vanishing gluon LCDA at low scales, $b_2(\mu_0)=0$, we find the charm quark contribution of $-1\%(-4\%)$
where the first number is for $\gamma^\ast\to \eta \gamma$ and the second one for  $\gamma^\ast\to \eta'\gamma$, respectively. 
For $b_2(\mu_0)=0.2$ we get $-5\%(-20\%)$ and for $b_2(\mu_0)=- 0.2$ one obtains $+3\%(+11\%)$. These numbers depend weakly on the
model of the light-quark LCDAs.  The $b$-quark contribution is much smaller in all cases.

\begin{figure}[ht]
\includegraphics[width=0.49\textwidth]{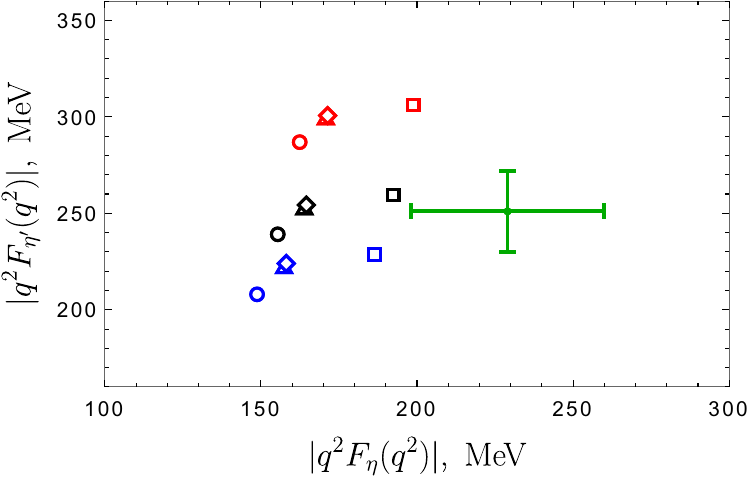}
\caption{
The $\gamma^\ast \to \eta'\gamma$ vs.  $\gamma^\ast \to \eta\gamma$ form factor at $q^2=112\, \text{GeV}^2$.
The experimental data point is from Ref.~\cite{BaBar:2006ash} with statistic and systematic errors added in 
quadrature. The three sets of points (from top to bottom) are obtained for the values of the gluon LCDA parameter \eqref{LCDA}
$b_2(\mu_0) = -0.2$ (red), $b_2(\mu_0) = 0$ (black) and $b_2(\mu_0) = 0.2$ (blue).
The results of the calculation with the four models of quark LCDAs specified in Table~\ref{table:1} (see also Fig.~\ref{figure:LCDAs})   
are shown with circles (I), triangles (II), diamonds (III) and squares (IV), respectively.    
}
\label{figure:FormFactors}
\end{figure}

Our final results for the four chosen models of the light-quark LCDAs and three different choices for the gluon LCDA are 
visualized in Fig.~\ref{figure:FormFactors}. The experimental value from Ref.~\cite{BaBar:2006ash} is also shown.
The size (and sign) of the gluon LCDA mainly affects the prediction for $\gamma^\ast \to \eta'\gamma$.
From the comparison with the data a vanishing (or small negative) gluon LCDA is preferred, although
the present experimental accuracy is not sufficient to draw definite conclusions. A much higher precision can be achieved at
Belle II  \cite{Belle-II:2018jsg}. 
The hierarchy of the results obtained using the four LCDA models (circles (I), triangles (II), diamonds (III) and squares (IV))    
is due mainly to the difference in the sum of the Gegenbauer coefficients
\begin{align}
\sigma^{(q)} = 1 + a_2^{(q)} +  a_4^{(q)} +  a_6^{(q)} +\ldots\,.
\end{align}
This can be expected as this sum enters the calculation of the form factors at the tree-level.
\begin{figure*}[t]
\includegraphics[width=0.42\textwidth]{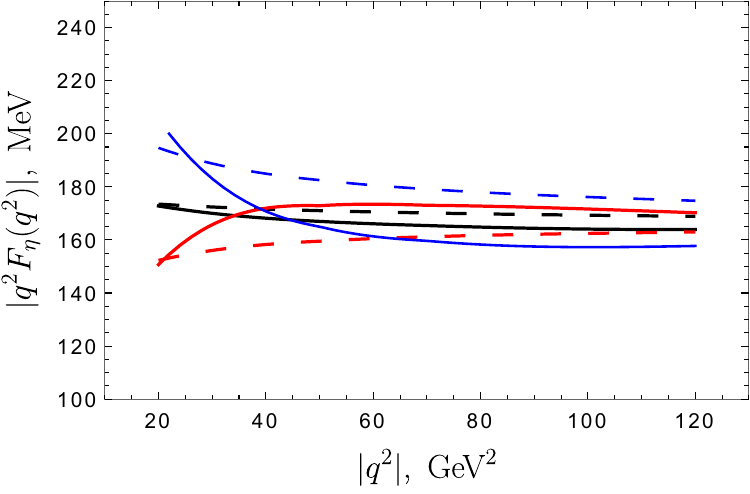}~~
\includegraphics[width=0.42\textwidth]{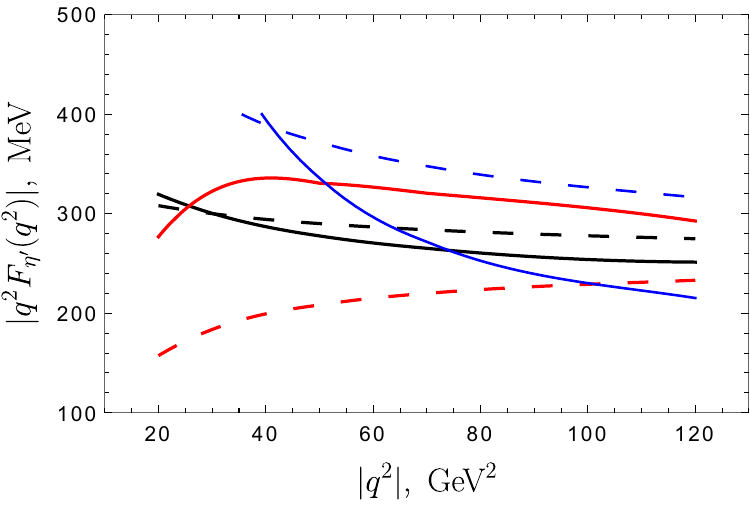}
\caption{Absolute values of the form factors $|q^2F_\eta(q^2)|$ (left panel) and  $|q^2F_{\eta'}(q^2)|$ (right panel) 
for time-like $q^2>0$ (solid curves) and space-like $q^2<0$ (dashed) photon virtualities.
The three pairs of curves correspond to the different choices of the gluon LCDA:
$b_2(\mu_0)=0$ (black), $b_2(\mu_0)= -0.2$ (red), and $b_2(\mu_0)=0.2$ (blue).   
}
\label{figure:EuMi}
\end{figure*}
For the models in Table~\ref{table:2} one obtains
\begin{align}
 \sigma^{(\ell)}(\mu_0) &= \{1.134, 1.228, 1.246, 1.399\}\,, 
\notag\\
 \sigma^{(s)}(\mu_0) &= \{1.115, 1.190, 1.208, 1.000\}\,,
\end{align} 
respectively.
These sums almost coincide for model~II and model~III, and the corresponding results for the 
form factors (triangles and diamonds) are very close to each other.
The results obtained using model I (circles) are roughly 5\% smaller for the both form factors as compared to models II, III.
Model~IV differs from the remaining ones by a much stronger $SU(3)$ violation which apparently leads to a decrease
of the $|F_{\eta'}/F_{\eta}|$ ratio. In this model (squares) the $\eta$ form factor comes out 17\% larger (compared to models II, III), 
but for  $\eta'$ the increase turns out to be  3\% only.
 
In Fig.~\ref{figure:EuMi} we compare the predictions for the form factors
for time-like $q^2>0$ (solid curves) and space-like $q^2<0$ (dashed) photon virtualities. 
The second quark LCDA model is used as an example.
As seen already from Table~\ref{table:3}, the time-like form factors acquire large phases. 
Their absolute values are, nevertherless, close to the space-like ones in a broad region of virtualities, 
provided the gluon LCDA is small. If the gluon LCDA is large, the difference between the time-like
and space-like $\eta'$ form factors can become very large as well. This difference may thus turn out to be 
a sensitive probe of the gluon contribution, provided the space-like form factors can be measured 
at sufficiently large $|q^2|$ to suppress power corrections.   

The subject of power corrections is complicated and a detailed discussion goes beyond the task of this paper.
In the space-like region, such corrections can be estimated using dispersion relations and duality, see 
Ref.~\cite{Agaev:2014wna} and references therein. Power corrections turn out to be, in general, large and negative, 
with the main effect being a suppression of the contribution of the end-point momentum fraction regions,
$u\to 0$ and $1-u\to 0$. A qualitatively similar suppression is expected also in the $k_T$-factorization approach 
going back to Ref.~\cite{Li:1992nu}. We are not aware of any solid approach to power-suppressed contributions in the 
time-like region, but, naively, the leading $1/q^2$  correction can be expected to be of opposite sign (i.e. positive).

The analysis in Ref.~\cite{Agaev:2014wna} is done to the NLO accuracy. 
It cannot be complemented by the NNLO contributions calculated in this work in any straightforward way, 
as higher-order perturbative corrections to transition form factors are increasingly sensitive to the end-point integration
regions due to Sudakov-type logarithms \cite{Braun:2020yib,Schoenleber:2022myb,Schoenleber:2024dvq}. Hence they are expected to
be affected even more strongly by nonperturbative effects. 
To illustrate this point, consider typical integrals contributing at the three consequent orders  
\begin{align}
2\int_{\frac12}^1 \!\! du\, T_{NS}^{(0)}(u)\, \phi^\ell_{\scriptscriptstyle II}(u) &= \phantom{-}6.922+0.400+0.005\,,
\notag\\
2\int_{\frac12}^1 \!\! du\, a_s T_{NS}^{(1)}(u)\, \phi^\ell_{\scriptscriptstyle II}(u) &= -1.104 + 0.310 + 0.101\,,
\notag\\
2 \int_{\frac12}^1 \!\! du\, a_s^2 T_{NS}^{(2)}(u)\, \phi^\ell_{\scriptscriptstyle II}(u) &= -0.972 +  0.275 + 0.197\,,
\label{end-point}
\end{align}
where the three numbers come from the integration regions $1/2 < u < 0.99$, $0.99 < u < 0.999$ and 
$0.999 < u < 1$, respectively, and
 we used the second LCDA model from Table~\ref{table:2} as an example,
$
 \phi^\ell_{\scriptscriptstyle II}(u) =  \Phi^{\ell}_{M,\text{\tiny Model II}}(u) /F_M^{\ell}.  
$
Since the  virtuality of the hard quark propagator in the Feynman diagrams for the CFs is $\sim (1-u)Q^2$,
for realistic values of $Q^2$  the last two regions are well beyond the applicability domain of the collinear expansion.
Such end-point contributuions are enhanced for the LCDAs with large higher-order Gegenbauer coefficients and become dramatic for models 
with non-standard $(\sim u(1-u))$ end-point behavior. For example, using a 
power-law ansatz in Eq.~\eqref{powerLCDA} as it stands, without the truncation at $n=8$, one obtains
\begin{align}
2 \int_{\frac12}^1 \!\! du\, a_s^2 T_{NS}^{(2)}(u)\, \phi^\ell_{p=0.532}(u) &= -0.967 + 0.495 +1.953\,.
\end{align}
Note that the contribution from the $u<0.99$ region is almost the same as in the last line in Eq.~\eqref{end-point}, which
means that the $n=8$ truncation has indeed high accuracy in the bulk of the momentum fraction region, but 
the end-point contributions become unphysically large.

These issues require a detailed study. Having them in mind, in this work we do not attempt the comparison of the results with the (existing) 
space like-data at $Q^2< 35$~GeV$^2$ and refer the reader to Refs.~\cite{Agaev:2014wna},\cite{Bali:2021qem} for the state-of-the-art
NLO analysis. An extension of these results to NNLO requires a calculation of the two-loop CFs with two different photon virtualities, 
cf.~\cite{Braun:2024srt}, which can be a separate large project.

\section{Summary}

We have carried out a NNLO calculation of the  time-like $\gamma^\ast \to \eta'\gamma$ and  $\gamma^\ast \to \eta\gamma$ form factors 
at the energy of the  $\Upsilon(4S)$ resonance, $s= q^2\simeq 112\, \text{GeV}^2$.
The same accuracy was achieved recently for the pion transition form factors \cite{Gao:2021iqq,Braun:2021grd}, 
but the present case is more complicated because of flavor-singlet contributions. In this work we have provided the last 
missing element for this calculation --- the three-loop anomalous dimension matrix for flavor-singlet axial-vector operators. 
Explicit expressions are given in the Appendix and also in the ancillary file (in Mathematica format).
Another new element in our analysis is the implementation of an NLO variable flavor number scheme for the charm quark contributions.

The results are presented in a model-independent form in Table~\ref{table:2}, and also in Fig.~\ref{figure:FormFactors} for several 
models of the light quark LCDA based on lattice QCD inputs. 
We find that the $\gamma^\ast \to \eta'\gamma$ form factor is sensitive to a possible admixture of gluons at a low scale, 
and the $F_{\eta'}/F_{\eta}$ ratio is also sensitive to the $SU(3)$ violations in the shape of the light quark LCDAs. 
The existing experimental data \cite{BaBar:2006ash} on  $\gamma^\ast \to \eta'\gamma$  
favor a small gluon LCDA, whereas there is some tension for $\gamma^\ast \to \eta\gamma$ calculated using quark LCDAs based on 
the existing lattice inputs \cite{Bali:2021qem,RQCD:2019osh,Cloet:2024vbv}. 

The accuracy of the NNLO calculation estimated by the remaining factorization scale dependence 
is very good, at the 1\% level, and the main remaining uncertainty is due to power-suppressed $1/q^2$ contributions.
A NNLO analysis of such corrections following the approach of Ref.~\cite{Agaev:2014wna} requires a calculation of 
the two-loop CFs with two different photon virtualities. This calculation is complicated but technically possible, 
cf.~\cite{Braun:2024srt}.


\section*{Acknowledgments}
This study was supported by Deutsche Forschungsgemeinschaft (DFG) through the Research Unit FOR 2926, ``Next Generation pQCD for
Hadron Structure: Preparing for the EIC'', project number 40824754. 
We thank Hua-Yu~Jiang for participation on an early stage of this project and S. Moch for providing us with 
the three-loop forward anomalous dimensions.


\appendix 
%
%

\section{Flavor-singlet anomalous dimensions}\label{App:A} 
In this Appendix, we present explicit expressions for the three-loop flavor-singlet AD matrix   
for QCD, $N_c=3$, in the form~\eqref{boldgamma}, \eqref{tildegamma}. 
The $\widetilde{\boldsymbol\gamma}_{00}$ element is a number~\cite{Larin:1993tq}:
\begin{align}
\widetilde{\boldsymbol\gamma}_{00} &= a^2_s\,16 n_f + a_s^3 \left( \frac{944}3 n_f -\frac{32}9 n_f^2\right).
\end{align}
The diagonal blocks $\widetilde{\boldsymbol\gamma}_{nn}$, $n\ge 2$, are $2\times 2$ matrices:
\begin{widetext}
\begin{align}
\widetilde{\boldsymbol\gamma}_{22} &=a_s
\begin{pmatrix}
\frac{100}9 & -\frac{1}3 n_f\\[2mm]
-\frac{40}9 & 18+\frac 43 n_f
\end{pmatrix}
+a_s^2
\left\{
\begin{pmatrix}
\frac{34450}{243} & 0\\[2mm]
-\frac{14590}{243}& \frac{447}{2}
\end{pmatrix}
+ n_f
\begin{pmatrix}
-\frac{745}{81} & -\frac{43}{324}\\[2mm]
-\frac{200}{81} & -\frac{1748}{81}
\end{pmatrix}
\right\}
\notag\\[2mm]
&\quad+
a_s^3\Biggl\{  \begin{pmatrix}
\frac{2200\zeta_3 }{81}+\frac{64486199}{26244} & 0\\[2mm]
-\frac{1600\zeta_3 }{81}-\frac{5535050}{6561} &\frac{28369}{8}
\end{pmatrix}
+n_f 
\begin{pmatrix}
 -\frac{4400\zeta_3 }{27}-\frac{1874755}{8748} & \frac{2560\zeta_3}{27}-\frac{1620961}{17496}     \\[2mm]
\frac{3200\zeta_3 }{27} +\frac{14155}{2187} & -\frac{14020\zeta_3}{27}-\frac{1460153}{17496}
\end{pmatrix}
+n_f^2\begin{pmatrix}
 -\frac{326}{81} & \frac{8869}{1458}       \\[2mm]
\frac{1330}{243} & -\frac{12305}{729}
\end{pmatrix}
\Biggr\},
\\[1mm]
\widetilde{\boldsymbol\gamma}_{44} &=a_s
\begin{pmatrix}
\frac{728}{45} & -\frac{2}{15} n_f \\[2mm]
-\frac{224}{45} & \frac{4}{3} n_f+\frac{156}{5}
\end{pmatrix}
+a_s^2
\left\{
\begin{pmatrix}
 \frac{662846}{3375} & 0 \\[2mm]
 -\frac{2307368}{30375} & \frac{126976}{375} 
\end{pmatrix}
+ n_f
\begin{pmatrix}
-\frac{151852}{10125} & \frac{22127}{20250} \\[2mm]
 -\frac{10528}{2025} & -\frac{375152}{10125} 
\end{pmatrix}
\right\}
\notag\\[2mm]
&\quad+
a_s^3\Biggl\{  \begin{pmatrix}
 \frac{11312 \zeta_3}{405}+\frac{559048023977}{164025000} & 0 \\[2mm]
 -\frac{21056 \zeta_3}{405}-\frac{43385387017}{41006250} & \frac{51647279}{9375} 
\end{pmatrix}
+n_f 
\begin{pmatrix}
-\frac{30016 \zeta_3}{135}-\frac{821093327}{2733750} & \frac{6476 \zeta_3}{135}-\frac{9414495251}{218700000} \\[2mm]
 \frac{3584 \zeta_3}{27}-\frac{104182414}{6834375} & -\frac{18976 \zeta_3}{27}-\frac{11563525657}{54675000} 
\end{pmatrix}
\notag\\[2mm]
&\quad
+n_f^2\begin{pmatrix}
 -\frac{743386}{151875} & \frac{2944457}{1518750} \\[2mm]
 \frac{135016}{30375} & -\frac{17837983}{911250} 
\end{pmatrix}
\Biggr\},
\\[3mm]
\widetilde{\boldsymbol\gamma}_{66} &=a_s
\begin{pmatrix}
\frac{2054}{105} & -\frac{1}{14}n_f \\[2mm]
 -\frac{36}{7} & \frac{1378}{35} +\frac43 n_f
\end{pmatrix}
+a_s^2
\Biggl\{
\begin{pmatrix}
 \frac{718751707}{3087000} & 0 \\[2mm]
 -\frac{36739}{420} & \frac{71321289}{171500} 
\end{pmatrix}
+ n_f
\begin{pmatrix}
 -\frac{103792931}{5556600} & \frac{210737}{211680} \\[2mm]
 -\frac{1766}{245} & -\frac{127437}{2744} 
\end{pmatrix}
\Biggr\}
\notag\\[2mm]
&\quad+
a_s^3\Biggl\{  \begin{pmatrix}
 \frac{185482 \zeta_3}{6615}+\frac{59388575317957639}{14702763600000} & 0 \\[2mm]
 -\frac{3672 \zeta_3}{49}-\frac{1286698543033}{1089093600} & \frac{11481783359213}{1680700000} 
\end{pmatrix}
\notag\\[2mm]
&\quad
+n_f 
\begin{pmatrix}
-\frac{116644 \zeta_3}{441}-\frac{344630882967529}{980184240000} & \frac{4237 \zeta_3}{147}-\frac{3330825303467}{130691232000} \\[2mm]
 \frac{960 \zeta_3}{7}-\frac{28588205419}{907578000} & -\frac{119753 \zeta_3}{147}-\frac{1587171468293}{5227649280} 
\end{pmatrix}
+n_f^2\begin{pmatrix}
  -\frac{19429628359}{3500658000} & \frac{27203147669}{32672808000} \\[2mm]
 \frac{286481}{102900} & -\frac{549102703}{25930800} 
\end{pmatrix}
\Biggr\}\,,
\\[3mm]
\widetilde{\boldsymbol\gamma}_{88} &=a_s
\begin{pmatrix}
 \frac{4180}{189} & -\frac{2}{45}n_f\\[2mm]
 -\frac{704}{135} & \frac{4763}{105} +\frac43 n_f
\end{pmatrix}
+a_s^2
\Biggl\{
\begin{pmatrix}
\frac{293323294583}{1125211500} & 0 \\[2mm]
 -\frac{3901134908}{40186125} & \frac{6600064319}{13891500} 
\end{pmatrix}
+ n_f
\begin{pmatrix}
  -\frac{287106517}{13395375} & \frac{1797557}{2187000} \\[2mm]
 -\frac{1119712}{127575} & -\frac{101892886}{1913625} 
\end{pmatrix}
\Biggr\}
\notag\\[2mm]
&\quad+
a_s^3\Biggl\{  \begin{pmatrix}
\frac{5012876 \zeta_3}{178605}+\frac{24226918396923088741}{5359157332200000} & 0 \\[2mm]
 -\frac{294712 \zeta_3}{25515}-\frac{12257491838900099}{76559390460000} & \frac{28836204088571687}{3675690900000} 
\end{pmatrix}
\notag\\[2mm]
&\quad
+n_f 
\begin{pmatrix}
-\frac{2527712 \zeta_3}{8505}-\frac{24860621334119521}{63799492050000} & \frac{32866 \zeta_3}{1701}-\frac{705201132190069}{40831674912000} \\[2mm]
 \frac{1408 \zeta_3}{81}-\frac{50854035270599}{9114213150000} & -\frac{1523848 \zeta_3}{1701}-\frac{1375186504728767}{3645685260000} 
\end{pmatrix}
+n_f^2\begin{pmatrix}
 -\frac{306389943931}{50634517500} & \frac{2502299986513}{6076142100000} \\[2mm]
 \frac{10522567}{80372250} & -\frac{161230169591}{7233502500} 
\end{pmatrix}
\Biggr\}.
\end{align}
The $\widetilde{\boldsymbol\gamma}_{n0}$ entries for $n\slashed{=}0$ are vectors:
\begin{align}
\widetilde{\boldsymbol\gamma}_{20} &=a^2_s 
\Biggl\{
\begin{pmatrix}
 \frac{260}{9} \\[2mm]
 -\frac{1400}{9} 
\end{pmatrix} 
+
n_f
\begin{pmatrix}
-\frac{16}{9} \\[2mm]
 -\frac{80}{9} 
\end{pmatrix} 
\Biggr\}
+a_s^3
\Biggl\{
\begin{pmatrix}
 \frac{49024}{81} \\[2mm]
 -\frac{235030}{81} 
\end{pmatrix} 
+
n_f
\begin{pmatrix}
-\frac{22175}{243} \\[2mm]
 \frac{58190}{243} 
\end{pmatrix} 
+
n_f^2
\begin{pmatrix}
\frac{65}{9} \\[2mm]
 \frac{220}{27} 
\end{pmatrix} 
\Biggr\},
\notag\\[2mm]
\widetilde{\boldsymbol\gamma}_{40} &=a^2_s 
\Biggl\{
\begin{pmatrix}
\frac{52}{9} \\[2mm]
 -\frac{11312}{45} 
\end{pmatrix} 
+
n_f
\begin{pmatrix}
 -\frac{56}{45} \\[2mm]
 -\frac{448}{45} 
\end{pmatrix} 
\Biggr\}
+a_s^3
\Biggl\{
\begin{pmatrix}
\frac{3911}{27} \\[2mm]
 -\frac{186096232}{30375} 
\end{pmatrix} 
+
n_f
\begin{pmatrix}
 -\frac{1216714}{30375} \\[2mm]
 \frac{9005668}{30375} 
\end{pmatrix} 
+
n_f^2
\begin{pmatrix}
 \frac{9388}{3375} \\[2mm]
 \frac{2576}{675} 
\end{pmatrix} 
\Biggr\},
\notag\\[2mm]
\widetilde{\boldsymbol\gamma}_{60} &=a^2_s 
\Biggl\{
\begin{pmatrix}
 -\frac{2054}{14175} \\[2mm]
 -\frac{4724}{15} 
\end{pmatrix} 
+
n_f
\begin{pmatrix}
 -\frac{2488}{2835} \\[2mm]
 -\frac{72}{7} 
\end{pmatrix} 
\Biggr\}
+a_s^3
\Biggl\{
\begin{pmatrix}
 \frac{281851388261}{7501410000} \\[2mm]
 -\frac{125803784333}{13891500} 
\end{pmatrix} 
+
n_f
\begin{pmatrix}
  -\frac{24453243641}{1000188000} \\[2mm]
 \frac{429338921}{1543500} 
\end{pmatrix} 
+
n_f^2
\begin{pmatrix} 
\frac{66779521}{50009400} \\[2mm]
 \frac{649}{735} 
\end{pmatrix} 
\Biggr\},
\notag\\[2mm]
\widetilde{\boldsymbol\gamma}_{80} &=a^2_s 
\Biggl\{
\begin{pmatrix}
-\frac{2717}{1323} \\[2mm]
 -\frac{1712392}{4725} 
\end{pmatrix} 
+
n_f
\begin{pmatrix}
 -\frac{614}{945} \\[2mm]
 -\frac{1408}{135} 
\end{pmatrix} 
\Biggr\}
+a_s^3
\Biggl\{
\begin{pmatrix}
 \frac{28739341349}{39382402500} \\[2mm]
 -\frac{66693211951949}{5626057500} 
\end{pmatrix} 
+
n_f
\begin{pmatrix}
 -\frac{65664119611}{3750705000} \\[2mm]
 \frac{22257179417}{93767625} 
\end{pmatrix} 
+
n_f^2
\begin{pmatrix} 
\frac{427531}{595350} \\[2mm]
 -\frac{52184}{42525} 
\end{pmatrix} 
\Biggr\}.
\end{align}
The remaining non-diagonal entries are again $2\times2$ matrices:
\begin{align}
\widetilde{\boldsymbol\gamma}_{42} &=a^2_s 
\Biggl\{
\begin{pmatrix}
\frac{8512}{243} & 0 \\[2mm]
 -\frac{33712}{405} & \frac{2156}{15} 
\end{pmatrix} 
+
n_f
\begin{pmatrix}
  -\frac{1036}{405} & \frac{49}{45} \\[2mm]
 -\frac{1568}{405} & -\frac{392}{405} 
\end{pmatrix} 
\Biggr\}
\notag\\[2mm]
&
+a_s^3
\Biggl\{
\begin{pmatrix}
 \frac{23599891}{36450} & 0 \\[2mm]
 -\frac{1043383901}{820125} & \frac{3601157}{1500} 
\end{pmatrix} 
+
n_f
\begin{pmatrix}
 -\frac{19974941}{182250} & \frac{8893087}{364500} \\[2mm]
 \frac{385238}{3375} & -\frac{63922607}{182250} 
\end{pmatrix} 
+
n_f^2
\begin{pmatrix} 
 \frac{62258}{30375} & -\frac{83279}{91125} \\[2mm]
 \frac{28616}{6075} & \frac{7154}{6075} 
\end{pmatrix} 
\Biggr\},
\notag\\[2mm]
\widetilde{\boldsymbol\gamma}_{62} &=a^2_s 
\Biggl\{
\begin{pmatrix}
\frac{34243}{2025} & 0 \\[2mm]
 -\frac{1316}{15} & \frac{18309}{125} 
\end{pmatrix} 
+
n_f
\begin{pmatrix}
  -\frac{79}{45} & \frac{511}{5400} \\[2mm]
 -4 & -1 
\end{pmatrix} 
\Biggr\}
\notag\\[2mm]
&\
+a_s^3
\Biggl\{
\begin{pmatrix}
\frac{208052194247}{714420000} & 0 \\[2mm]
 -\frac{1552505147}{992250} & \frac{6469961437}{2450000} 
\end{pmatrix} 
+
n_f
\begin{pmatrix}
  -\frac{48465845261}{857304000} & \frac{30434567717}{2857680000} \\[2mm]
 \frac{34499593}{330750} & -\frac{321089627}{882000} 
\end{pmatrix} 
+
n_f^2
\begin{pmatrix} 
 \frac{4993943}{4762800} & -\frac{4391567}{57153600} \\[2mm]
 \frac{163}{70} & \frac{163}{280} 
\end{pmatrix} 
\Biggr\},
\notag\\[2mm]
\widetilde{\boldsymbol\gamma}_{82} &=a^2_s 
\Biggl\{
\begin{pmatrix}
\frac{47072}{5103} & 0 \\[2mm]
 -\frac{1761188}{18225} & \frac{542113}{3150} 
\end{pmatrix} 
+
n_f
\begin{pmatrix}
-\frac{524}{405} & -\frac{749}{6075} \\[2mm]
 -\frac{4928}{1215} & -\frac{1232}{1215} 
\end{pmatrix} 
\Biggr\}
\notag\\[2mm]
&\quad
+a_s^3
\Biggl\{
\begin{pmatrix}
 \frac{4068118032481}{25317258750} & 0 \\[2mm]
 -\frac{13600382918837}{7233502500} & \frac{1879385623621}{555660000} 
\end{pmatrix} 
+
n_f
\begin{pmatrix}
 -\frac{83347630159}{2411167500} & \frac{165674357677}{19289340000} \\[2mm]
 \frac{11016032456}{120558375} & -\frac{259972344731}{602791875} 
\end{pmatrix} 
+
n_f^2
\begin{pmatrix} 
\frac{642487}{1148175} & \frac{49253}{918540} \\[2mm]
 \frac{49808}{54675} & \frac{12452}{54675} 
\end{pmatrix} 
\Biggr\},
\notag\\[2mm]
\widetilde{\boldsymbol\gamma}_{64} &=a^2_s 
\Biggl\{
\begin{pmatrix}
\frac{2208998}{70875} & 0 \\[2mm]
 -\frac{28732}{375} & \frac{93291}{875} 
\end{pmatrix} 
+
n_f
\begin{pmatrix}
-\frac{2992}{1575} & \frac{12859}{13500} \\[2mm]
 -\frac{88}{35} & -\frac{44}{175} 
\end{pmatrix} 
\Biggr\}
\notag\\[2mm]
&\quad
+a_s^3
\Biggl\{
\begin{pmatrix}
 \frac{21898269506047}{37507050000} & 0 \\[2mm]
 -\frac{464483888297}{347287500} & \frac{1493390417}{857500} 
\end{pmatrix} 
+
n_f
\begin{pmatrix}
 -\frac{6712204620473}{75014100000} & \frac{188005248607}{16669800000} \\[2mm]
 \frac{2195028209}{23152500} & -\frac{3227274787}{12862500} 
\end{pmatrix} 
+
n_f^2
\begin{pmatrix} 
 \frac{131277949}{83349000} & -\frac{155192939}{312558750} \\[2mm]
 \frac{14927}{3675} & \frac{14927}{36750} 
\end{pmatrix} 
\Biggr\},
\notag\\[2mm]
\widetilde{\boldsymbol\gamma}_{84} &=a^2_s 
\Biggl\{
\begin{pmatrix}
\frac{80220734}{4465125} & 0 \\[2mm]
 -\frac{43984952}{637875} & \frac{26809123}{308700} 
\end{pmatrix} 
+
n_f
\begin{pmatrix}
 -\frac{102476}{70875} & \frac{1050401}{2976750} \\[2mm]
 -\frac{15488}{6075} & -\frac{7744}{30375} 
\end{pmatrix} 
\Biggr\}
+a_s^3
\Biggl\{
\begin{pmatrix}
 \frac{547136173376507}{1772208112500} & 0 \\[2mm]
 -\frac{89274580744877}{63293146875} & \frac{382689123131843}{272273400000} 
\end{pmatrix} 
\notag\\[2mm]
&\quad
+
n_f
\begin{pmatrix}
  -\frac{878219057771}{16878172500} & \frac{6612995092253}{2362944150000} \\[2mm]
 \frac{1617306158842}{21097715625} & -\frac{4391083811213}{21097715625} 
\end{pmatrix} 
+
n_f^2
\begin{pmatrix} 
 \frac{203845367}{200930625} & -\frac{751254031}{8037225000} \\[2mm]
 \frac{4063664}{1913625} & \frac{2031832}{9568125} 
\end{pmatrix} 
\Biggr\},
\notag\\[2mm]
\widetilde{\boldsymbol\gamma}_{86} &=a^2_s 
\Biggl\{
\begin{pmatrix}
\frac{2552663}{92610} & 0 \\[2mm]
 -\frac{3447268}{46305} & \frac{154253}{1715} 
\end{pmatrix} 
+
n_f
\begin{pmatrix}
 -\frac{1985}{1323} & \frac{79913}{123480} \\[2mm]
 -\frac{352}{189} & -\frac{44}{441} 
\end{pmatrix} 
\Biggr\}
+a_s^3
\Biggl\{
\begin{pmatrix}
  \frac{58135173181189}{110270727000} & 0 \\[2mm]
 -\frac{10975089254723}{7876480500} & \frac{779605161401}{518616000} 
\end{pmatrix} 
\notag\\[2mm]
&\quad
+
n_f
\begin{pmatrix}
  -\frac{22482165445907}{294055272000} & \frac{510346526357}{84015792000} \\[2mm]
 \frac{8858934271}{105019740} & -\frac{30767605351571}{147027636000} 
\end{pmatrix} 
+
n_f^2
\begin{pmatrix} 
 \frac{2190169}{1666980} & -\frac{10108073}{38896200} \\[2mm]
 \frac{229636}{59535} & \frac{57409}{277830} 
\end{pmatrix} 
\Biggr\}.
\end{align}
\end{widetext}
These expressions present our main result. The corresponding flavor-nonsinglet AD matrices 
are available from Refs.~\cite{Moch:2014sna,Braun:2021tzi}. They are collected, for completeness, in the provided ancillary file.


\bibliographystyle{apsrev}
\bibliography{references}

\end{document}